\documentclass[zpreprint,zbstepj]{zeus_paper}
\usepackage[english]{babel}
\usepackage{lineno}
\usepackage{units}
\newcommand{\PYTHIA}{\textsc{Pythia}\xspace}
\newcommand{\pTb}{\ensuremath{p_{T}^{b}}\xspace}

\newcommand{\pTcjet}{\ensuremath{p_{T}^{c\text{-jet}}}\xspace}
\newcommand{\pT}{\ensuremath{p_{T}}\xspace}
\newcommand{\pTjet}{\ensuremath{p_{T}^{\text{jet}}}\xspace}
\newcommand{\etajet}{\ensuremath{\eta^{\text{jet}}}\xspace}
\newcommand{\etaqjet}{\ensuremath{\eta^{q-\text{jet}}}\xspace}
\newcommand{\mvtx}{\ensuremath{m_{\text{vtx}}}\xspace}
\newcommand{\ntrk}{\ensuremath{n_{\text{trk}}}\xspace}
\newcommand{\xgamma}{\ensuremath{x_{\gamma}^{\text{jet}}}\xspace}
\newcommand{\Qsq}{\ensuremath{{Q^{2}}}\xspace}
\newcommand{\bbbar}{\ensuremath{b\bbar}\xspace}
\newcommand{\ccbar}{\ensuremath{c\cbar}\xspace}
\newcommand{\dif}{\text{d}}
\newcommand{\diffpt}{\ensuremath{\dif\sigma/\dif\pTjet}\xspace}
\newcommand{\diffeta}{\ensuremath{\dif\sigma/\dif\etajet}\xspace}
\newcommand{\diffptb}{\ensuremath{\dif\sigma_{b}/\dif\pTjet}\xspace}
\newcommand{\diffetab}{\ensuremath{\dif\sigma_{b}/\dif\etajet}\xspace}
\newcommand{\diffptc}{\ensuremath{\dif\sigma_{c}/\dif\pTjet}\xspace}
\newcommand{\diffetac}{\ensuremath{\dif\sigma_{c}/\dif\etajet}\xspace}
\newcommand{\diffnloptb}{\ensuremath{\dif\sigma_{b}^{\text{NLO}}/\dif\pTjet\otimes C^{b}_{\text{had}}}\xspace}
\newcommand{\diffnloetab}{\ensuremath{\dif\sigma_{b}^{\text{NLO}}/\dif\etajet\otimes C^{b}_{\text{had}}}\xspace}
\newcommand{\diffnloptc}{\ensuremath{\dif\sigma_{c}^{\text{NLO}}/\dif\pTjet\otimes C^{c}_{\text{had}}}\xspace}
\newcommand{\diffnloetac}{\ensuremath{\dif\sigma_{c}^{\text{NLO}}/\dif\etajet\otimes C^{c}_{\text{had}}}\xspace}
\newcommand{\cbhad}{\ensuremath{C^{b}_{\text{had}}}\xspace}
\newcommand{\cchad}{\ensuremath{C^{c}_{\text{had}}}\xspace}

\newcommand{\ndf}{\ensuremath{\text{ndf}}\xspace}
\newcommand{\Zdetdesc}{%
A detailed description of the ZEUS detector can be found 
elsewhere~\cite{zeus:1993:bluebook}. A brief outline of the 
components that are most relevant for this analysis is given
below.\xspace}
\newcommand{\Zctdmvddesc}[1]{%
In the kinematic range of the analysis, charged particles were tracked
in the central tracking detector (CTD)~\citeCTD, and the microvertex
detector (MVD)~\citeMVD. These components operated in a magnetic
field of $1.43\Tesla$ provided by a thin superconducting solenoid. The
CTD consisted of 72~cylindrical drift chamber layers, organised in
nine superlayers covering the polar-angle#1 region
\mbox{$15^\circ<\theta<164^\circ$}. The MVD consisted of a barrel
(BMVD) and a forward (FMVD) section with three cylindrical layers and
four vertical planes of single-sided silicon strip sensors in the BMVD
and FMVD respectively. The BMVD provided polar-angle coverage for
tracks with three measurements from $30^\circ$ to $150^\circ$. The
FMVD extended the polar-angle coverage in the forward region to
$7^\circ$. After alignment the single hit resolution of the BMVD was
$\rm 25\,\mu m$ and the impact parameter resolution of the CTD-BMVD system
for high momentum tracks was $\rm 100\,\mu m$.}
\newcommand{\Zsttdesc}[1]{%
The STT consisted of 48 sectors of two different sizes. Each sector
contained 192 (small sector) or 264 (large sector) straws of diameter
7.5 mm arranged into 3 layers. The sectors were trapezoidal in shape
and each subtended an azimuthal angle of $60^{\circ}$ -- 6 sectors
formed a so-called superlayer. A particle passing through the complete
detector traversed 8 superlayers, which were rotated around the beam
direction at angles of $30^{\circ}$ or $15^{\circ}$ to each other. The STT
covered the polar-angle region $5^{\circ}<\theta<23^{\circ}$.
}
\newcommand{\Zcaldesc}{%
The high-resolution uranium--scintillator calorimeter (CAL)~\citeCAL consisted 
of three parts: the forward (FCAL), the barrel (BCAL) and the rear (RCAL)
calorimeters. Each part was subdivided transversely into towers and
longitudinally into one electromagnetic section (EMC) and either one (in RCAL)
or two (in BCAL and FCAL) hadronic sections (HAC). The smallest subdivision of
the calorimeter is called a cell.  The CAL energy resolutions, as measured under
test-beam conditions, are $\sigma(E)/E=0.18/\sqrt{E}$ for electrons and
$\sigma(E)/E=0.35/\sqrt{E}$ for hadrons ($E$ in $\Gev$).}
\newcommand{\Zlumidesc}[1]{%
The luminosity was measured using the Bethe-Heitler reaction
$ep\,\rightarrow\, e\gamma p$ by a luminosity detector which consisted
of independent lead-scintillator calorimeter\citePCAL and a magnetic
spectrometer\citeSPECTRO systems. The fractional systematic
uncertainty on the measured luminosity was #1.}
\chardef\usc=95
\chardef\til=126
\catcode`\@=11 % @ signs are now treated as letters
\DeclareRobustCommand\xdotspace{\futurelet\@let@token\@xdotspace}
\def\@xdotspace{%
  \ifx\@let@token.\else
  \ifx\@let@token\bgroup.\else
  \ifx\@let@token\egroup.\else
  \ifx\@let@token\/.\else
  \ifx\@let@token\ .\else
  \ifx\@let@token~.\else
  \ifx\@let@token!.\else
  \ifx\@let@token,.\else
  \ifx\@let@token:.\else
  \ifx\@let@token;.\else
  \ifx\@let@token?.\else
  \ifx\@let@token/.\else
  \ifx\@let@token'.\else
  \ifx\@let@token).\else
  \ifx\@let@token-.\else
  \ifx\@let@token\@xobeysp.\else
  \ifx\@let@token\space.\else
  \ifx\@let@token\@sptoken.\else
   .\space
   \fi\fi\fi\fi\fi\fi\fi\fi\fi\fi\fi\fi\fi\fi\fi\fi\fi\fi}
\catcode`\@=12 % @ signs are no longer letters

\newcommand{\stru}[2]{%
   \relax\ifmmode\hbox{\vrule height#1 depth#2 width0pt}%
   \else\vrule height#1 depth#2 width0pt\fi}
\newcommand{\Ronum}[1]{\uppercase\expandafter{\romannumeral#1}}
\newcommand{\ronum}[1]{\expandafter{\romannumeral#1}}
\DeclareRobustCommand{\LaTeXZ}{%
  \LaTeX\kern-.05em4\kern-.1em
  {\raisebox{-0.2ex}{$\scriptstyle\text{ZEUS}$}}\xspace}
\DeclareMathAlphabet{\mathbf}{OT1}{cmr}{bx}{sl}
\newcommand{\eVdist}{\kern-0.06667em}

\newcommand{\Gev}{{\text{Ge}\eVdist\text{V\/}}}

\newcommand{\gev}{{\,\text{Ge}\eVdist\text{V\/}}}

\newcommand{\pbi}{\,\text{pb}^{-1}}

\newcommand{\Tesla}{\,\text{T}}

\newcommand{\slashfrac}[2]{%
  \raisebox{0.5ex}{\ensuremath #1}\kern-0.12em/\kern-0.08em
  \raisebox{-.8ex}{\ensuremath #2}}
\newcommand{\sqr}[3]{%
    {\vcenter{\hrule height.#3ex\hbox{\vrule width.#2ex height#1ex
     \kern#1ex\vrule width.#3ex}\hrule height.#2ex}}}

\newcommand{\widebar}[1]{%
   \mkern1.5mu\overline{\mkern-1.5mu#1\mkern-1.mu}\mkern1.mu}
\catcode`\@=11 % @ signs are now treated as letters
\newcommand{\parenbar}{\mathpalette\p@renb@r}
\def\p@renb@r#1#2{\vbox{%
  \ifx#1\scriptscriptstyle \dimen@.7em\dimen@ii.2em\else
  \ifx#1\scriptstyle \dimen@.8em\dimen@ii.25em\else
  \dimen@1em\dimen@ii.4em\fi\fi \offinterlineskip
  \ialign{\hfill##\hfill\cr
    \vbox{\hrule width\dimen@ii}\cr
    \noalign{\vskip-.3ex}%
    \hbox to\dimen@{$\mathchar300\hfil\mathchar301$}\cr
    \noalign{\vskip-.3ex}%
    $#1#2$\cr}}}
\catcode`\@=12 % @ signs are no longer letters

\newcommand{\cbar}{\widebar{c}}
\newcommand{\bbar}{\widebar{b}}

\newcommand{\IP}{{\rm I$\kern-0.01667em$P}\xspace}

\mathchardef\qsm=63
\mathchardef\pls=43
\mathchardef\mns=512
\mathchardef\plm=518
\mathchardef\eql=61
\mathchardef\smallleft=300
\mathchardef\smallright=301
\mathchardef\les=316
\mathchardef\gre=318
\mathchardef\leq=532
\mathchardef\grq=533
\catcode`\@=11 % @ signs are now treated as letters
\newcounter{pict@width}
\newcounter{pict@height}
\newlength{\pict@scale}
\newcommand{\psfigadd}[4]{%
\setcounter{pict@width}{1*\ratio{#2+\pict@scale/2}{\pict@scale}}
\setcounter{pict@height}{1*\ratio{#3+\pict@scale/2}{\pict@scale}}
\setlength{\unitlength}{\pict@scale}
\hbox to #2{\hspace{-\fill}\begin{picture}(\thepict@width,\thepict@height)
\put(0,0){\psfig{figure=#1,width=#2,height=#3,clip=}}
\SetScale{0.283466457}
\SetWidth{1.763889}
{#4}
\end{picture}}
}
\newcounter{pict@widthfst}
\newcounter{pict@widthscd}
\newcounter{pict@widthtot}
\newcommand{\psfigaddtwo}[7]{%
\setcounter{pict@widthfst}{1*\ratio{#2+\pict@scale/2}{\pict@scale}}
\setcounter{pict@widthscd}{1*\ratio{#2+#4+\pict@scale/2}{\pict@scale}}
\setcounter{pict@widthtot}{1*\ratio{#2+#4+#6+\pict@scale/2}{\pict@scale}}
\setcounter{pict@height}{1*\ratio{#3+\pict@scale/2}{\pict@scale}}
\setlength{\unitlength}{\pict@scale}
\hbox{\hspace{-\fill}\begin{picture}(\thepict@widthtot,\thepict@height)
\put(0,0){\psfig{figure=#1,width=#2,height=#3,clip=}}
\put(\thepict@widthscd,0){\psfig{figure=#5,width=#6,height=#3,clip=}}
\SetScale{0.283466457}
\SetWidth{1.763889}
{#7}
\end{picture}}
}
\newcommand{\psfigror}[4]{%
\setcounter{pict@width}{1*\ratio{#2+\pict@scale/2}{\pict@scale}}
\setcounter{pict@height}{1*\ratio{#3+\pict@scale/2}{\pict@scale}}
\setlength{\unitlength}{\pict@scale}
\hbox{\begin{picture}(\thepict@width,\thepict@height)
\put(0,\thepict@height){\psfig{figure=#1,width=#3,height=#2,clip=,angle=270}}
\SetScale{0.283466457}
\SetWidth{1.763889}
{#4}
\end{picture}}
}
\newcommand{\psfigrol}[4]{%
\setcounter{pict@width}{1*\ratio{#2+\pict@scale/2}{\pict@scale}}
\setcounter{pict@height}{1*\ratio{#3+\pict@scale/2}{\pict@scale}}
\setlength{\unitlength}{\pict@scale}
\hbox{\begin{picture}(\thepict@width,\thepict@height)
\put(0,0){\psfig{figure=#1,width=#3,height=#2,clip=,angle=90}}
\SetScale{0.283466457}
\SetWidth{1.763889}
{#4}
\end{picture}}
}
\catcode`\@=12 % @ signs are no longer letters
\newlength\listtextwidth

\catcode`\@=11 % @ signs are now treated as letters
\newlength{\@tabfninsert}
\newlength{\@tabfnwidth}
\newcommand{\tabfootnote}[2]{%
  \setlength{\@tabfninsert}{0.8em}
  \setlength{\@tabfnwidth}{\textwidth}
  \addtolength{\@tabfnwidth}{-\@tabfninsert}
  \addtolength{\@tabfnwidth}{-0.4em}
  \noindent\makebox[\@tabfninsert][r]{\footnotesize$^{#1}$\hfil}\hfill%
  \parbox[t]{\@tabfnwidth}{\footnotesize #2\hfill}}
\catcode`\@=12 % @ signs are no longer letters

\renewcommand{\Zctdmvddesc}[1]{%
In the kinematic range of the analysis, charged particles were tracked
in the central tracking detector (CTD)~\citeCTD and the microvertex
detector (MVD)~\citeMVD. These components operated in a magnetic
field of $1.43\Tesla$ provided by a thin superconducting solenoid. The
CTD consisted of 72~cylindrical drift-chamber layers, organised in nine
superlayers covering the polar-angle#1 region
\mbox{$15^\circ<\theta<164^\circ$}.
The MVD silicon tracker consisted of a barrel (BMVD) and a forward
(FMVD) section. The BMVD contained three layers and provided
polar-angle coverage for tracks from $30^\circ$ to
$150^\circ$. The four-layer FMVD extended the polar-angle coverage in
the forward region to $7^\circ$. After alignment, the single-hit
resolution of the MVD was $\unit[24]{\mu m}$. The transverse distance of closest
approach (DCA) to the nominal vertex in $X$--$Y$ was measured to have
a resolution, averaged over the azimuthal angle, of $\unit[(46 \oplus 122 /
p_{T})]{\mu m}$, with $p_{T}$ in \Gev.  For CTD-MVD tracks that pass
through all nine CTD superlayers, the momentum resolution was
$\sigma(p_{T})/p_{T} = 0.0029 p_{T} \oplus 0.0081 \oplus
0.0012/p_{T}$, with $p_{T}$ in \Gev.}

\newcommand{\Zcoosys}{\footnote{The ZEUS coordinate system is a
    right-handed Cartesian system, with the $Z$ axis pointing in the
    proton beam direction, referred to as the ``forward direction'',
    and the $X$ axis pointing left towards the centre of HERA.  The
    coordinate origin is at the centre of the CTD. The
    pseudorapidity is defined as
    $\eta=-\ln\left(\tan\frac{\theta}{2}\right)$, where the polar
    angle, $\theta$, is measured with respect to the proton beam
    direction. The azimuthal angle, $\phi$, is measured with respect
    to the $X$ axis.}}

\renewcommand{\Zcaldesc}{%
  The high-resolution uranium--scintillator calorimeter (CAL)~\citeCAL
  consisted of three parts: the forward (FCAL), the barrel (BCAL) and
  the rear (RCAL) calorimeters. Each part was subdivided transversely
  into towers and longitudinally into one electromagnetic section
  (EMC) and either one (in RCAL) or two (in BCAL and FCAL) hadronic
  sections (HAC). The smallest subdivision of the calorimeter was
  called a cell.  The CAL energy resolutions, as measured under
  test-beam conditions, were $\sigma(E)/E=0.18/\sqrt{E}$ for electrons
  and $\sigma(E)/E=0.35/\sqrt{E}$ for hadrons, with $E$ in $\Gev$.}

\renewcommand{\Zlumidesc}[1]{%
  The luminosity was measured using the Bethe-Heitler reaction
  $ep\,\rightarrow\, e\gamma p$ by a luminosity detector which
  consisted of independent lead--scintillator calorimeter\citePCAL and
  magnetic spectrometer\cite{nim:a565:572} systems. The fractional
  systematic uncertainty on the measured luminosity was #1.}

\newcommand{\stat}{\ensuremath{\text{stat.}}}
\newcommand{\syst}{\ensuremath{\text{syst.}}}

\def\citeCTD{{\cite{%
nim:a279:290,*npps:b32:181,*nim:a338:254%
}}\xspace}
\def\citeMVD{{\cite{%
nim:a581:656%
}}\xspace}
\def\citeCAL{{\cite{%
nim:a309:77,*nim:a309:101,*nim:a321:356,*nim:a336:23%
}}\xspace}
\def\cite6mT{{\cite{%
thesis:gosau:2007%
}}\xspace}
\def\citePCAL{{\cite{%
desy-92-066,*zfp:c63:391,*acpp:b32:2025%
}}\xspace}

\begin{document}

\prepnum{DESY--11--067}

\title{Measurement of heavy-quark jet photoproduction at HERA}

\author{ZEUS Collaboration}
\date{28 April 2011}

\abstract{Photoproduction of beauty and charm quarks in events with at
  least two jets has been measured with the ZEUS detector at HERA
  using an integrated luminosity of $\unit[133]{\pbi}$. The fractions
  of jets containing $b$ and $c$ quarks were extracted using the
  invariant mass of charged tracks associated with secondary vertices
  and the decay-length significance of these vertices. Differential
  cross sections as a function of jet transverse momentum, \pTjet, and
  pseudorapidity, \etajet, were measured.
  The data are compared with previous measurements and are well described
  by next-to-leading-order QCD predictions.}

\makezeustitle
\def\3{\ss}
\pagenumbering{Roman}

\begin{center}
  {\Large The ZEUS Collaboration}
\end{center}

{\small
  
%  members:

        {\raggedright
H.~Abramowicz$^{45, ai}$, 
I.~Abt$^{35}$, 
L.~Adamczyk$^{13}$, 
M.~Adamus$^{54}$, 
R.~Aggarwal$^{7, d}$, 
S.~Antonelli$^{4}$, 
P.~Antonioli$^{3}$, 
A.~Antonov$^{33}$, 
M.~Arneodo$^{50}$, 
V.~Aushev$^{26, 27, ab}$
% ...
        }

\newpage

%       institutes:

\makebox[3em]{$^{1}$}
\begin{minipage}[t]{14cm}
{\it Argonne National Laboratory, Argonne, Illinois 60439-4815, USA}~$^{A}$

\end{minipage}\\
\makebox[3em]{$^{2}$}
\begin{minipage}[t]{14cm}
{\it Andrews University, Berrien Springs, Michigan 49104-0380, USA}

\end{minipage}\\
\makebox[3em]{$^{3}$}
\begin{minipage}[t]{14cm}
{\it INFN Bologna, Bologna, Italy}~$^{B}$

\end{minipage}\\
\makebox[3em]{$^{4}$}
\begin{minipage}[t]{14cm}
{\it University and INFN Bologna, Bologna, Italy}~$^{B}$

\end{minipage}\\
\makebox[3em]{$^{5}$}
\begin{minipage}[t]{14cm}
{\it Physikalisches Institut der Universit\"at Bonn,
Bonn, Germany}~$^{C}$

\end{minipage}\\
\makebox[3em]{$^{6}$}
\begin{minipage}[t]{14cm}
{\it H.H.~Wills Physics Laboratory, University of Bristol,
Bristol, United Kingdom}~$^{D}$

\end{minipage}\\
\makebox[3em]{$^{7}$}
\begin{minipage}[t]{14cm}
{\it Panjab University, Department of Physics, Chandigarh, India}

\end{minipage}\\
\makebox[3em]{$^{8}$}
\begin{minipage}[t]{14cm}
{\it Calabria University,
Physics Department and INFN, Cosenza, Italy}~$^{B}$

\end{minipage}\\
\makebox[3em]{$^{9}$}
\begin{minipage}[t]{14cm}
{\it Institute for Universe and Elementary Particles, Chonnam National University,\\
Kwangju, South Korea}

\end{minipage}\\
\makebox[3em]{$^{10}$}
\begin{minipage}[t]{14cm}
{\it Jabatan Fizik, Universiti Malaya, 50603 Kuala Lumpur, Malaysia}~$^{E}$

\end{minipage}\\
\makebox[3em]{$^{11}$}
\begin{minipage}[t]{14cm}
{\it Nevis Laboratories, Columbia University, Irvington on Hudson,
New York 10027, USA}~$^{F}$

\end{minipage}\\
\makebox[3em]{$^{12}$}
\begin{minipage}[t]{14cm}
{\it The Henryk Niewodniczanski Institute of Nuclear Physics, Polish Academy of \\
Sciences, Cracow, Poland}~$^{G}$

\end{minipage}\\
\makebox[3em]{$^{13}$}
\begin{minipage}[t]{14cm}
{\it Faculty of Physics and Applied Computer Science, AGH-University of Science and \\
Technology, Cracow, Poland}~$^{H}$

\end{minipage}\\
\makebox[3em]{$^{14}$}
\begin{minipage}[t]{14cm}
{\it Department of Physics, Jagellonian University, Cracow, Poland}

\end{minipage}\\
\makebox[3em]{$^{15}$}
\begin{minipage}[t]{14cm}
{\it Deutsches Elektronen-Synchrotron DESY, Hamburg, Germany}

\end{minipage}\\
\makebox[3em]{$^{16}$}
\begin{minipage}[t]{14cm}
{\it Deutsches Elektronen-Synchrotron DESY, Zeuthen, Germany}

\end{minipage}\\
\makebox[3em]{$^{17}$}
\begin{minipage}[t]{14cm}
{\it INFN Florence, Florence, Italy}~$^{B}$

\end{minipage}\\
\makebox[3em]{$^{18}$}
\begin{minipage}[t]{14cm}
{\it University and INFN Florence, Florence, Italy}~$^{B}$

\end{minipage}\\
\makebox[3em]{$^{19}$}
\begin{minipage}[t]{14cm}
{\it Fakult\"at f\"ur Physik der Universit\"at Freiburg i.Br.,
Freiburg i.Br., Germany}

\end{minipage}\\
\makebox[3em]{$^{20}$}
\begin{minipage}[t]{14cm}
{\it School of Physics and Astronomy, University of Glasgow,
Glasgow, United Kingdom}~$^{D}$

\end{minipage}\\
\makebox[3em]{$^{21}$}
\begin{minipage}[t]{14cm}
{\it Department of Engineering in Management and Finance, Univ. of
the Aegean, Chios, Greece}

\end{minipage}\\
\makebox[3em]{$^{22}$}
\begin{minipage}[t]{14cm}
{\it Hamburg University, Institute of Experimental Physics, Hamburg,
Germany}~$^{I}$

\end{minipage}\\
\makebox[3em]{$^{23}$}
\begin{minipage}[t]{14cm}
{\it Imperial College London, High Energy Nuclear Physics Group,
London, United Kingdom}~$^{D}$

\end{minipage}\\
\makebox[3em]{$^{24}$}
\begin{minipage}[t]{14cm}
{\it Institute of Particle and Nuclear Studies, KEK,
Tsukuba, Japan}~$^{J}$

\end{minipage}\\
\makebox[3em]{$^{25}$}
\begin{minipage}[t]{14cm}
{\it Institute of Physics and Technology of Ministry of Education and
Science of Kazakhstan, Almaty, Kazakhstan}

\end{minipage}\\
\makebox[3em]{$^{26}$}
\begin{minipage}[t]{14cm}
{\it Institute for Nuclear Research, National Academy of Sciences, Kyiv, Ukraine}

\end{minipage}\\
\makebox[3em]{$^{27}$}
\begin{minipage}[t]{14cm}
{\it Department of Nuclear Physics, National Taras Shevchenko University of Kyiv, Kyiv, Ukraine}

\end{minipage}\\
\makebox[3em]{$^{28}$}
\begin{minipage}[t]{14cm}
{\it Kyungpook National University, Center for High Energy Physics, Daegu,
South Korea}~$^{K}$

\end{minipage}\\
\makebox[3em]{$^{29}$}
\begin{minipage}[t]{14cm}
{\it Institut de Physique Nucl\'{e}aire, Universit\'{e} Catholique de Louvain, Louvain-la-Neuve,\\
Belgium}~$^{L}$

\end{minipage}\\
\makebox[3em]{$^{30}$}
\begin{minipage}[t]{14cm}
{\it Departamento de F\'{\i}sica Te\'orica, Universidad Aut\'onoma
de Madrid, Madrid, Spain}~$^{M}$

\end{minipage}\\
\makebox[3em]{$^{31}$}
\begin{minipage}[t]{14cm}
{\it Department of Physics, McGill University,
Montr\'eal, Qu\'ebec, Canada H3A 2T8}~$^{N}$

\end{minipage}\\
\makebox[3em]{$^{32}$}
\begin{minipage}[t]{14cm}
{\it Meiji Gakuin University, Faculty of General Education,
Yokohama, Japan}~$^{J}$

\end{minipage}\\
\makebox[3em]{$^{33}$}
\begin{minipage}[t]{14cm}
{\it Moscow Engineering Physics Institute, Moscow, Russia}~$^{O}$

\end{minipage}\\
\makebox[3em]{$^{34}$}
\begin{minipage}[t]{14cm}
{\it Moscow State University, Institute of Nuclear Physics,
Moscow, Russia}~$^{P}$

\end{minipage}\\
\makebox[3em]{$^{35}$}
\begin{minipage}[t]{14cm}
{\it Max-Planck-Institut f\"ur Physik, M\"unchen, Germany}

\end{minipage}\\
\makebox[3em]{$^{36}$}
\begin{minipage}[t]{14cm}
{\it NIKHEF and University of Amsterdam, Amsterdam, Netherlands}~$^{Q}$

\end{minipage}\\
\makebox[3em]{$^{37}$}
\begin{minipage}[t]{14cm}
{\it Physics Department, Ohio State University,
Columbus, Ohio 43210, USA}~$^{A}$

\end{minipage}\\
\makebox[3em]{$^{38}$}
\begin{minipage}[t]{14cm}
{\it Department of Physics, University of Oxford,
Oxford, United Kingdom}~$^{D}$

\end{minipage}\\
\makebox[3em]{$^{39}$}
\begin{minipage}[t]{14cm}
{\it INFN Padova, Padova, Italy}~$^{B}$

\end{minipage}\\
\makebox[3em]{$^{40}$}
\begin{minipage}[t]{14cm}
{\it Dipartimento di Fisica dell' Universit\`a and INFN,
Padova, Italy}~$^{B}$

\end{minipage}\\
\makebox[3em]{$^{41}$}
\begin{minipage}[t]{14cm}
{\it Department of Physics, Pennsylvania State University, University Park,\\
Pennsylvania 16802, USA}~$^{F}$

\end{minipage}\\
\makebox[3em]{$^{42}$}
\begin{minipage}[t]{14cm}
{\it Polytechnic University, Sagamihara, Japan}~$^{J}$

\end{minipage}\\
\makebox[3em]{$^{43}$}
\begin{minipage}[t]{14cm}
{\it Dipartimento di Fisica, Universit\`a 'La Sapienza' and INFN,
Rome, Italy}~$^{B}$

\end{minipage}\\
\makebox[3em]{$^{44}$}
\begin{minipage}[t]{14cm}
{\it Rutherford Appleton Laboratory, Chilton, Didcot, Oxon,
United Kingdom}~$^{D}$

\end{minipage}\\
\makebox[3em]{$^{45}$}
\begin{minipage}[t]{14cm}
{\it Raymond and Beverly Sackler Faculty of Exact Sciences, School of Physics, \\
Tel Aviv University, Tel Aviv, Israel}~$^{R}$

\end{minipage}\\
\makebox[3em]{$^{46}$}
\begin{minipage}[t]{14cm}
{\it Department of Physics, Tokyo Institute of Technology,
Tokyo, Japan}~$^{J}$

\end{minipage}\\
\makebox[3em]{$^{47}$}
\begin{minipage}[t]{14cm}
{\it Department of Physics, University of Tokyo,
Tokyo, Japan}~$^{J}$

\end{minipage}\\
\makebox[3em]{$^{48}$}
\begin{minipage}[t]{14cm}
{\it Tokyo Metropolitan University, Department of Physics,
Tokyo, Japan}~$^{J}$

\end{minipage}\\
\makebox[3em]{$^{49}$}
\begin{minipage}[t]{14cm}
{\it Universit\`a di Torino and INFN, Torino, Italy}~$^{B}$

\end{minipage}\\
\makebox[3em]{$^{50}$}
\begin{minipage}[t]{14cm}
{\it Universit\`a del Piemonte Orientale, Novara, and INFN, Torino,
Italy}~$^{B}$

\end{minipage}\\
\makebox[3em]{$^{51}$}
\begin{minipage}[t]{14cm}
{\it Department of Physics, University of Toronto, Toronto, Ontario,
Canada M5S 1A7}~$^{N}$

\end{minipage}\\
\makebox[3em]{$^{52}$}
\begin{minipage}[t]{14cm}
{\it Physics and Astronomy Department, University College London,
London, United Kingdom}~$^{D}$

\end{minipage}\\
\makebox[3em]{$^{53}$}
\begin{minipage}[t]{14cm}
{\it Faculty of Physics, University of Warsaw, Warsaw, Poland}

\end{minipage}\\
\makebox[3em]{$^{54}$}
\begin{minipage}[t]{14cm}
{\it Institute for Nuclear Studies, Warsaw, Poland}

\end{minipage}\\
\makebox[3em]{$^{55}$}
\begin{minipage}[t]{14cm}
{\it Department of Particle Physics and Astrophysics, Weizmann
Institute, Rehovot, Israel}

\end{minipage}\\
\makebox[3em]{$^{56}$}
\begin{minipage}[t]{14cm}
{\it Department of Physics, University of Wisconsin, Madison,
Wisconsin 53706, USA}~$^{A}$

\end{minipage}\\
\makebox[3em]{$^{57}$}
\begin{minipage}[t]{14cm}
{\it Department of Physics, York University, Ontario, Canada M3J
1P3}~$^{N}$

\end{minipage}\\
\vspace{30em} \pagebreak[4]

%  references concerning institutes;

\makebox[3ex]{$^{ A}$}
\begin{minipage}[t]{14cm}
 supported by the US Department of Energy\
\end{minipage}\\
\makebox[3ex]{$^{ B}$}
\begin{minipage}[t]{14cm}
 supported by the Italian National Institute for Nuclear Physics (INFN) \
\end{minipage}\\
\makebox[3ex]{$^{ C}$}
\begin{minipage}[t]{14cm}
 supported by the German Federal Ministry for Education and Research (BMBF), under
 contract No. 05 H09PDF\
\end{minipage}\\
\makebox[3ex]{$^{ D}$}
\begin{minipage}[t]{14cm}
 supported by the Science and Technology Facilities Council, UK\
\end{minipage}\\
\makebox[3ex]{$^{ E}$}
\begin{minipage}[t]{14cm}
 supported by an FRGS grant from the Malaysian government\
\end{minipage}\\
\makebox[3ex]{$^{ F}$}
\begin{minipage}[t]{14cm}
 supported by the US National Science Foundation. Any opinion,
 findings and conclusions or recommendations expressed in this material
 are those of the authors and do not necessarily reflect the views of the
 National Science Foundation.\
\end{minipage}\\
\makebox[3ex]{$^{ G}$}
\begin{minipage}[t]{14cm}
 supported by the Polish Ministry of Science and Higher Education as a scientific project No.
 DPN/N188/DESY/2009\
\end{minipage}\\
\makebox[3ex]{$^{ H}$}
\begin{minipage}[t]{14cm}
 supported by the Polish Ministry of Science and Higher Education
 as a scientific project (2009-2010)\
\end{minipage}\\
\makebox[3ex]{$^{ I}$}
\begin{minipage}[t]{14cm}
 supported by the German Federal Ministry for Education and Research (BMBF), under
 contract No. 05h09GUF, and the SFB 676 of the Deutsche Forschungsgemeinschaft (DFG) \
\end{minipage}\\
\makebox[3ex]{$^{ J}$}
\begin{minipage}[t]{14cm}
 supported by the Japanese Ministry of Education, Culture, Sports, Science and Technology
 (MEXT) and its grants for Scientific Research\
\end{minipage}\\
\makebox[3ex]{$^{ K}$}
\begin{minipage}[t]{14cm}
 supported by the Korean Ministry of Education and Korea Science and Engineering
 Foundation\
\end{minipage}\\
\makebox[3ex]{$^{ L}$}
\begin{minipage}[t]{14cm}
 supported by FNRS and its associated funds (IISN and FRIA) and by an Inter-University
 Attraction Poles Programme subsidised by the Belgian Federal Science Policy Office\
\end{minipage}\\
\makebox[3ex]{$^{ M}$}
\begin{minipage}[t]{14cm}
 supported by the Spanish Ministry of Education and Science through funds provided by
 CICYT\
\end{minipage}\\
\makebox[3ex]{$^{ N}$}
\begin{minipage}[t]{14cm}
 supported by the Natural Sciences and Engineering Research Council of Canada (NSERC) \
\end{minipage}\\
\makebox[3ex]{$^{ O}$}
\begin{minipage}[t]{14cm}
 partially supported by the German Federal Ministry for Education and Research (BMBF)\
\end{minipage}\\
\makebox[3ex]{$^{ P}$}
\begin{minipage}[t]{14cm}
 supported by RF Presidential grant N 41-42.2010.2 for the Leading
 Scientific Schools and by the Russian Ministry of Education and Science through its
 grant for Scientific Research on High Energy Physics\
\end{minipage}\\
\makebox[3ex]{$^{ Q}$}
\begin{minipage}[t]{14cm}
 supported by the Netherlands Foundation for Research on Matter (FOM)\
\end{minipage}\\
\makebox[3ex]{$^{ R}$}
\begin{minipage}[t]{14cm}
 supported by the Israel Science Foundation\
\end{minipage}\\
\vspace{30em} \pagebreak[4]

%  references concerning mebers;

\makebox[3ex]{$^{ a}$}
\begin{minipage}[t]{14cm}
also affiliated with University College London,
 United Kingdom\
\end{minipage}\\
\makebox[3ex]{$^{ b}$}
\begin{minipage}[t]{14cm}
now at University of Salerno, Italy\
\end{minipage}\\
\makebox[3ex]{$^{ c}$}
\begin{minipage}[t]{14cm}
now at Queen Mary University of London, United Kingdom\
\end{minipage}\\
\makebox[3ex]{$^{ d}$}
\begin{minipage}[t]{14cm}
also funded by Max Planck Institute for Physics, Munich, Germany\
\end{minipage}\\
\makebox[3ex]{$^{ e}$}
\begin{minipage}[t]{14cm}
also Senior Alexander von Humboldt Research Fellow at Hamburg University,
 Institute of Experimental Physics, Hamburg, Germany\
\end{minipage}\\
\makebox[3ex]{$^{ f}$}
\begin{minipage}[t]{14cm}
also at Cracow University of Technology, Faculty of Physics,
 Mathemathics and Applied Computer Science, Poland\
\end{minipage}\\
\makebox[3ex]{$^{ g}$}
\begin{minipage}[t]{14cm}
supported by the research grant No. 1 P03B 04529 (2005-2008)\
\end{minipage}\\
\makebox[3ex]{$^{ h}$}
\begin{minipage}[t]{14cm}
now at Rockefeller University, New York, NY
 10065, USA\
\end{minipage}\\
\makebox[3ex]{$^{ i}$}
\begin{minipage}[t]{14cm}
now at DESY group FS-CFEL-1\
\end{minipage}\\
\makebox[3ex]{$^{ j}$}
\begin{minipage}[t]{14cm}
now at Institute of High Energy Physics, Beijing, China\
\end{minipage}\\
\makebox[3ex]{$^{ k}$}
\begin{minipage}[t]{14cm}
now at DESY group FEB, Hamburg, Germany\
\end{minipage}\\
\makebox[3ex]{$^{ l}$}
\begin{minipage}[t]{14cm}
also at Moscow State University, Russia\
\end{minipage}\\
\makebox[3ex]{$^{ m}$}
\begin{minipage}[t]{14cm}
now at University of Liverpool, United Kingdom\
\end{minipage}\\
\makebox[3ex]{$^{ n}$}
\begin{minipage}[t]{14cm}
now at CERN, Geneva, Switzerland\
\end{minipage}\\
\makebox[3ex]{$^{ o}$}
\begin{minipage}[t]{14cm}
also affiliated with Universtiy College London, UK\
\end{minipage}\\
\makebox[3ex]{$^{ p}$}
\begin{minipage}[t]{14cm}
now at Goldman Sachs, London, UK\
\end{minipage}\\
\makebox[3ex]{$^{ q}$}
\begin{minipage}[t]{14cm}
also at Institute of Theoretical and Experimental Physics, Moscow, Russia\
\end{minipage}\\
\makebox[3ex]{$^{ r}$}
\begin{minipage}[t]{14cm}
also at INP, Cracow, Poland\
\end{minipage}\\
\makebox[3ex]{$^{ s}$}
\begin{minipage}[t]{14cm}
also at FPACS, AGH-UST, Cracow, Poland\
\end{minipage}\\
\makebox[3ex]{$^{ t}$}
\begin{minipage}[t]{14cm}
partially supported by Warsaw University, Poland\
\end{minipage}\\
\makebox[3ex]{$^{ u}$}
\begin{minipage}[t]{14cm}
now at Istituto Nucleare di Fisica Nazionale (INFN), Pisa, Italy\
\end{minipage}\\
\makebox[3ex]{$^{ v}$}
\begin{minipage}[t]{14cm}
now at Haase Energie Technik AG, Neum\"unster, Germany\
\end{minipage}\\
\makebox[3ex]{$^{ w}$}
\begin{minipage}[t]{14cm}
now at Department of Physics, University of Bonn, Germany\
\end{minipage}\\
\makebox[3ex]{$^{ x}$}
\begin{minipage}[t]{14cm}
now at Biodiversit\"at und Klimaforschungszentrum (BiK-F), Frankfurt, Germany\
\end{minipage}\\
\makebox[3ex]{$^{ y}$}
\begin{minipage}[t]{14cm}
also affiliated with DESY, Germany\
\end{minipage}\\
\makebox[3ex]{$^{ z}$}
\begin{minipage}[t]{14cm}
also at University of Tokyo, Japan\
\end{minipage}\\
\makebox[3ex]{$^{\dagger}$}
\begin{minipage}[t]{14cm}
 deceased \
\end{minipage}\\
\makebox[3ex]{$^{aa}$}
\begin{minipage}[t]{14cm}
now at Kobe University, Japan\
\end{minipage}\\
\makebox[3ex]{$^{ab}$}
\begin{minipage}[t]{14cm}
supported by DESY, Germany\
\end{minipage}\\
\makebox[3ex]{$^{ac}$}
\begin{minipage}[t]{14cm}
member of National Technical University of Ukraine, Kyiv Polytechnic Institute, Kyiv,
 Ukraine\
\end{minipage}\\
\makebox[3ex]{$^{ad}$}
\begin{minipage}[t]{14cm}
member of National University of Kyiv - Mohyla Academy, Kyiv, Ukraine\
\end{minipage}\\
\makebox[3ex]{$^{ae}$}
\begin{minipage}[t]{14cm}
supported by the Bogolyubov Institute for Theoretical Physics of the National
 Academy of Sciences, Ukraine\
\end{minipage}\\
\makebox[3ex]{$^{af}$}
\begin{minipage}[t]{14cm}
STFC Advanced Fellow\
\end{minipage}\\
\makebox[3ex]{$^{ag}$}
\begin{minipage}[t]{14cm}
nee Korcsak-Gorzo\
\end{minipage}\\
\makebox[3ex]{$^{ah}$}
\begin{minipage}[t]{14cm}
This material was based on work supported by the
 National Science Foundation, while working at the Foundation.\
\end{minipage}\\
\makebox[3ex]{$^{ai}$}
\begin{minipage}[t]{14cm}
also at Max Planck Institute for Physics, Munich, Germany, External Scientific Member\
\end{minipage}\\
\makebox[3ex]{$^{aj}$}
\begin{minipage}[t]{14cm}
now at Tokyo Metropolitan University, Japan\
\end{minipage}\\
\makebox[3ex]{$^{ak}$}
\begin{minipage}[t]{14cm}
now at Nihon Institute of Medical Science, Japan\
\end{minipage}\\
\makebox[3ex]{$^{al}$}
\begin{minipage}[t]{14cm}
now at Osaka University, Osaka, Japan\
\end{minipage}\\
\makebox[3ex]{$^{am}$}
\begin{minipage}[t]{14cm}
also at \L\'{o}d\'{z} University, Poland\
\end{minipage}\\
\makebox[3ex]{$^{an}$}
\begin{minipage}[t]{14cm}
member of \L\'{o}d\'{z} University, Poland\
\end{minipage}\\
\makebox[3ex]{$^{ao}$}
\begin{minipage}[t]{14cm}
now at Lund University, Lund, Sweden\
\end{minipage}\\
\makebox[3ex]{$^{ap}$}
\begin{minipage}[t]{14cm}
also at University of Podlasie, Siedlce, Poland\
\end{minipage}\\

}

%------------------------------------------------------------------------------
%       Text
%------------------------------------------------------------------------------
\pagenumbering{arabic} 
\pagestyle{plain}
% ----------------------------------------------------------------------------
%       Introduction
% ----------------------------------------------------------------------------
\section{Introduction}
\label{sec-int}

The study of beauty and charm production in $ep$ collisions
constitutes a rigorous test of perturbative Quantum Chromodynamics
(QCD) since the heavy-quark masses provide a hard scale that allows
perturbative calculations. At leading order, boson-gluon fusion (BGF),
$\gamma g\rightarrow q\bar q$ with $q\in\{b,c\}$, is the dominant
process for heavy-quark production at HERA. When the negative squared
four-momentum exchanged at the electron vertex, \Qsq, is small, the
process can be treated as photoproduction, in which a quasi-real
photon emitted by the incoming electron interacts with the proton. For
heavy-quark transverse momenta larger than or comparable to the quark
mass, next-to-leading-order (NLO) QCD calculations in which the
massive quarks are generated in the hard
sub-process\cite{np:b454:3f,*pl:b348:633} are expected to provide
reliable predictions for the photoproduction cross sections.

Beauty and charm photoproduction has been measured using several
different methods by both the ZEUS and H1 collaborations.  In most of
the previous measurements of beauty photoproduction at HERA, the cross
section was determined using semileptonic decays into
muons\cite{pr:d70:012008b,Chekanov:2008zz,Chekanov:2008tx,epj:c41:453b}
or electrons\cite{epj:c18:625,pr:d78:072001a}. In the muon analyses,
the fraction of leptons originating from beauty was determined by
using the large transverse momentum of the muon relative to the axis
of the associated jet, $p_{T}^{\text{rel}}$, and/or exploiting the
impact parameter of the muons. In the more recent electron
analysis\cite{epj:c18:625}, several variables, sensitive to both
electron identification as well as to semileptonic decays, were
combined in a likelihood-ratio test function in order to extract the
beauty and charm content.  The H1 collaboration has published an
inclusive measurement of beauty- and charm-quark photoproduction using
a method based on the impact parameter of tracks to the primary
vertex\cite{epj:c47:597}.  The other published charm or beauty
photoproduction
measurements\cite{Breitweg:1998yt,Chekanov2005492,epj:c44:351,epj:c50:299,pl:b565:87,epj:c50:251,pl:b621:56}
used either meson tags or a combination of lepton and meson tags.  In
all of the above analyses reasonable agreement between the measurement
and the theory prediction was found.

The aim of this measurement is to test perturbative QCD with high
precision. For this purpose, the long lifetimes of the weakly decaying
$b$ and $c$ hadrons as well as their large masses were exploited.  The
measurement relies on the reconstruction of decay vertices with the
ZEUS silicon microvertex detector (MVD)\citeMVD. Two discriminating
variables were used: the significance of the reconstructed decay
length and the invariant mass of the charged tracks associated with the
decay vertex (secondary vertex). The measurement was kept fully
inclusive, leading to a reduced uncertainty due to branching fractions
and a substantial increase in statistics compared to exclusive
analyses.  The high statistics also allowed the kinematic region of
the measurement to be extended to high values of the transverse jet
momentum, \pTjet.

% ----------------------------------------------------------------------------
%       Experimental set-up
% ----------------------------------------------------------------------------
\section{Experimental set-up}
\label{sec-exp}

The analysis was performed with data corresponding to an integrated
luminosity of $\unit[133]{\pbi}$ which were taken during 2005.
Electrons at an energy of $E_{e}=\unit[27.5]{\Gev}$ collided with
protons at $E_{p}=\unit[920]{\Gev}$, yielding a centre-of-mass energy
of $\unit[318]{\Gev}$.

% ----------------------------------------------------------------------------
%       General detector blabla
% ----------------------------------------------------------------------------
\Zdetdesc

% ----------------------------------------------------------------------------
%       CTD+MVD description, footnote on coordinate system is the argument
% ----------------------------------------------------------------------------
\Zctdmvddesc\Zcoosys

% ----------------------------------------------------------------------------
%       CAL description straight and simple
% ----------------------------------------------------------------------------
\Zcaldesc

% ----------------------------------------------------------------------------
%       Muon system description. You may want it or not. No references for it
% ----------------------------------------------------------------------------
%\Zmuondesc

% ----------------------------------------------------------------------------
%       BAC description complete with reference
% ----------------------------------------------------------------------------
%\Zbacdesc

% ----------------------------------------------------------------------------
%       LUMI, give actual luminosity uncertainty for your sample as argument
% ----------------------------------------------------------------------------
\Zlumidesc{1.8~\%}

% ----------------------------------------------------------------------------
%       Monte Carlo section
% ----------------------------------------------------------------------------
\section{Monte Carlo simulation}
\label{sec:montecarlo}

Monte Carlo (MC) samples of
beauty, charm and light-flavour events generated with
\PYTHIA~6.2\cite{cpc:135:238,*epj:c17:137,*hep-ph-0108264} were used
to evaluate the detector acceptance and to provide the predictions of
the signal and background distributions.

The production of \bbbar{} and \ccbar{} pairs was simulated following the standard
\PYTHIA{} prescription, using leading-order matrix elements combined with
parton showering. The following
subprocesses\cite{thesis:vschoenb:2010} were generated:
\begin{itemize}
\item direct and resolved photoproduction with leading-order massive
  matrix elements. In the direct-photon process, the quasi-real photon
  enters directly in the hard interaction, while in the
  resolved-photon process, the photon acts as a source of light
  partons which take part in the hard interaction.  The $b$-quark and
  $c$-quark masses were set to $4.75\gev$ and $1.5\gev$, respectively;
\item $b$-quark and $c$-quark excitation, i.e.\ the contribution to
  the leading-order massless matrix elements of
  $b$ and $c$ quarks from initial-state photon or gluon splitting.
\end{itemize}

The light-quark predictions
were taken from a simulation of both direct and non-direct inclusive
photoproduction with leading-order matrix elements in the massless
scheme. This sample also includes final-state gluon splitting into
\bbbar and \ccbar pairs, which is treated as part of the signal.

The CTEQ4L\cite{pr:d55:1280} and CTEQ5L\cite{epj:c12:375u} proton
parton distribution functions (PDFs) were used for the light-flavour
and heavy-flavour samples, respectively. The GRV-G
LO\cite{pr:d46:1973, *pr:d45:3986} photon PDF was used for all samples.

The lifetimes of the $B^{\pm}$, $B^{0}$ and $B_{s}$ mesons were
corrected from the default \PYTHIA values to reflect the
world averages\cite{pdg2008}.

The generated events were passed through a full simulation of the ZEUS
detector based on \textsc{Geant}~3.21\cite{tech:cern-dd-ee-84-1}. The
final MC events had to fulfil the same trigger requirements
and pass the same reconstruction programme as the data.

% ----------------------------------------------------------------------------
%      Data Selection
% ----------------------------------------------------------------------------
\section{Data selection and event reconstruction}
\label{data-sel}

A three-level trigger system was used to select events
online\cite{zeus:1993:bluebook,nim:a580:1257,proc:chep:1992:222}. At
the third level, jets were reconstructed using the energies and
positions in the CAL. Events with at least two jets with transverse
momentum in excess of $\unit[4.5]{\Gev}$ within $|\eta|<\unit[2.5]{}$
were selected.

The tracking efficiency at the first-level trigger (FLT) as well as
the efficiency of the dijet third-level trigger (TLT) were lowered
in the detector simulation such that they reproduced the efficiencies
as measured in the data. The trigger efficiencies were
$\approx\unit[86]{\%}$ for the FLT and \unit[76--100]{\%} for the TLT, depending
on the transverse momentum of the jets, with an average of about
\unit[90]{\%}. The average corrections amounted
to $\approx\unit[7.7]{\%}$ for the FLT and $\approx\unit[3.7]{\%}$ for
the TLT.

The hadronic system was reconstructed from energy-flow objects
(EFOs)\cite{thesis:briskin:1998} combining track and calorimeter
information, corrected for energy loss in the dead material. Each EFO,
$i$, was assigned a reconstructed four-momentum
$(p^{i}_{X},p^{i}_{Y},p^{i}_{Z},E^{i})$, assuming the pion mass.  Jets
were reconstructed from EFOs using a $k_{T}$ clustering
algorithm\cite{np:b406:187n} in the longitudinally invariant
mode\cite{pr:d48:3160}. The $E$-recombination scheme, which produces
massive jets whose four-momenta are the sum of the four-momenta of the
clustered objects, was used.

At least two jets with $|\etajet|<2.5$ and $\pTjet>\unit[7(6)]{\Gev}$
for the highest (second highest) energetic jet were required.
Only events with a well reconstructed primary vertex with
$|Z_{\text{vtx}}|<\unit[30]{cm}$ were selected.

In order to remove background from deep inelastic scattering (DIS),
events were rejected in which a scattered-electron candidate was found
in the calorimeter with energy $E'_{e} > 5\gev$ and $y_{e} < 0.9$,
with $y_{e}= 1- \frac{E'_{e}}{2E_{e}}\left( 1-\cos{\theta'_{e}}
\right)$, where $\theta'_{e}$ is the polar angle of the outgoing
electron. The event inelasticity, $y$, was reconstructed from the
hadronic final state using the Jacquet-Blondel
method\cite{proc:epfacility:1979:391} with $y_{\text{JB}} =
\sum_{i}(E^{i}-p_{Z}^{i})/2E_{e}$, where the sum runs over all the
EFOs. A cut $0.2<y_{\text{JB}}<0.8$ was used to remove residual DIS
events and non-$ep$ interactions. These requirements correspond to an
effective cut of $\Qsq \lesssim \unit[1]{\Gev^{2}}$ with a median of
$\Qsq \approx\unit[10^{-3}]{\Gev^{2}}$, as estimated from simulations.

In order to reconstruct secondary vertices related to $b$- and
$c$-hadron decays, tracks were selected if:
\begin{itemize}
\item $\pT>\unit[0.5]{\Gev}$;
\item the number of superlayers in the CTD $\ge 3$;
\item the total number of hits\footnote{Each MVD
    layer provided two coordinate measurements.} in the MVD $\ge 4$.
\end{itemize}

The tracks were associated with one of the two highest energetic jets if
they fulfilled $\Delta R =
\sqrt{(\eta^{\text{trk}}-\eta^{\text{jet}})^{2}+(\phi^{\text{trk}}-\phi^{\text{jet}})^{2}}
< 1$. If two or more of such tracks were associated with the selected
jet, a candidate vertex was fitted from the selected tracks using a
deterministic annealing
filter\cite{RosePhysRevLett.65.945,*Rose726788,*Didierjean2010188}. This
fit provided the vertex position including its error matrix as well as
the invariant mass, \mvtx, of the charged tracks associated with the
reconstructed vertex.  Vertices with $\chi^{2}/\ndf < 6$, a distance
from the interaction point within $\unit[1]{cm}$ in the $X$--$Y$
plane and $\pm \unit[30]{cm}$ in the $Z$ direction, and $\unit[0.8]{}\le
\mvtx < \unit[7.5]{\Gev}$ were retained for further analysis.

Only those secondary vertices that were associated with one of the two
jets with the highest $\pTjet$ were considered,
since these jets were most likely to correspond to heavy-quark
jets. The associated jet was required to be reconstructed within the
central part of the detector with $\unit[-1.6]{}\le\etajet<\unit[1.4]{}$. 

% ----------------------------------------------------------------------------
%     Extraction of the beauty fraction
% ----------------------------------------------------------------------------
\section{Extraction of the heavy-flavour cross sections}
\label{bb-extr}

Using the secondary vertices associated with jets, the decay length,
$d$, was defined as the distance in $X$--$Y$ between the secondary
vertex and the interaction point\footnote{%
  In the $X$--$Y$ plane, the interaction point was defined as the
  centre of the beam ellipse, determined using the average primary
  vertex position for groups of a few thousand events, taking into
  account the difference in angle between the beam direction and the
  $Z$ direction. The $Z$ coordinate was taken as the $Z$ position of
  the primary vertex of the event.}, projected onto the jet axis
in the $X$--$Y$ plane.

The decay-length significance, $S$, was defined as $d/\delta d$, where
$\delta d$ is the uncertainty on $d$. The sign of the decay length was
assigned using the axis of the jet to which the vertex is associated:
if the decay-length vector was in the same hemisphere as the jet axis,
a positive sign was assigned to it; otherwise the sign of the decay
length was negative. Negative decay lengths, which originate from
secondary vertices reconstructed on the wrong side of the interaction
point with respect to the direction of the associated jets, are
unphysical and caused by detector resolution effects. A small
correction\cite{thesis:vschoenb:2010} to the MC decay-length
distribution was applied in order to reproduce the negative
decay-length data: $\unit[5]{\%}$ of the tracks in the central region
were smeared and an additional smearing was applied to tracks in the
tails of the decay-length distribution.

\begin{figure}[p]
\centering
\includegraphics[scale=0.75]{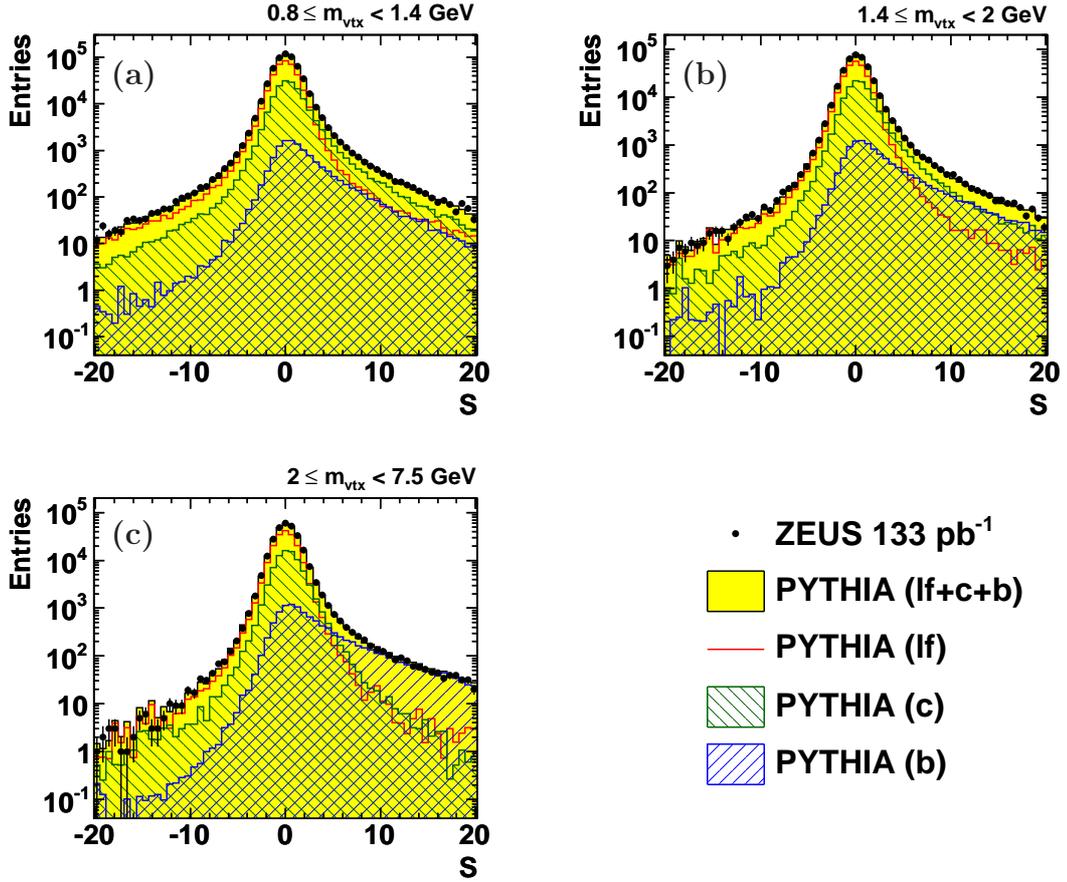}
\put(-13.2,11.5){\bf (a)}
\put(-5.7,11.5){\bf (b)}
\put(-13.2,5.4){\bf (c)}
\caption{Distributions of decay-length significance, $S$, for (a)
  $\unit[0.8]{} \le \mvtx < \unit[1.4]{\gev}$, (b) $\unit[1.4]{} \le
  \mvtx < \unit[2]{\gev}$ and (c) $\unit[2]{}\le \mvtx <
  \unit[7.5]{\gev}$.  The data are compared to the total PYTHIA MC
  distributions as well as the contributions from the beauty, charm
  and light-flavour MC subsamples. All samples were normalised
  according to the scaling factors obtained from the fit (see
  Section~\ref{bb-results}).  }
\label{fig_dl}
\end{figure}

The shape of the decay-length significance distribution together with
the secondary-vertex mass distribution, \mvtx, is used to extract the
beauty and charm content. The invariant mass of the tracks fitted to
the secondary vertex provides a distinguishing variable for jets from
$b$ and $c$ quarks, reflecting the different masses of the $b$ and $c$
hadrons. Figure~\ref{fig_dl} shows the decay-length significance, $S$,
divided into the three mass bins $\unit[0.8]{} \le \mvtx <
\unit[1.4]{\Gev}$, $\unit[1.4]{} \le \mvtx < \unit[2]{\Gev}$ and
$\unit[2]{}\le \mvtx < \unit[7.5]{\Gev}$. The MC simulation provides a
good description of the data in all three bins and an almost pure
beauty region can be obtained at high significances in the bin
$\unit[2]{}\le \mvtx < \unit[7.5]{\Gev}$.

In order to minimise the effect of the light-flavour contribution, the
contents of the negative bins of the significance distribution,
$N(S^{-})$, were subtracted from the contents of the corresponding
positive bins, $N(S^{+})$, yielding a subtracted decay-length
significance distribution. An additional advantage of this subtraction
is that symmetric systematic effects, which might arise from
discrepancies between the data and the MC, are removed.

In order further to reduce the uncertainty due to remaining
differences between data and MC in the core region of the significance
distribution, a cut of $|S|>3$ was applied. As a consistency check this
cut was varied in order to estimate the uncertainty due to the MC
modelling of the low $|S|$ region; effects smaller than
$\unit[1]{\%}$ on the beauty results and $\unit[3]{\%}$ on the charm
results were found.

After all selection cuts, a sample of 70\,433 jets with associated
secondary vertices remained.

\begin{figure}[p]
\centering
 \includegraphics[scale=0.75]{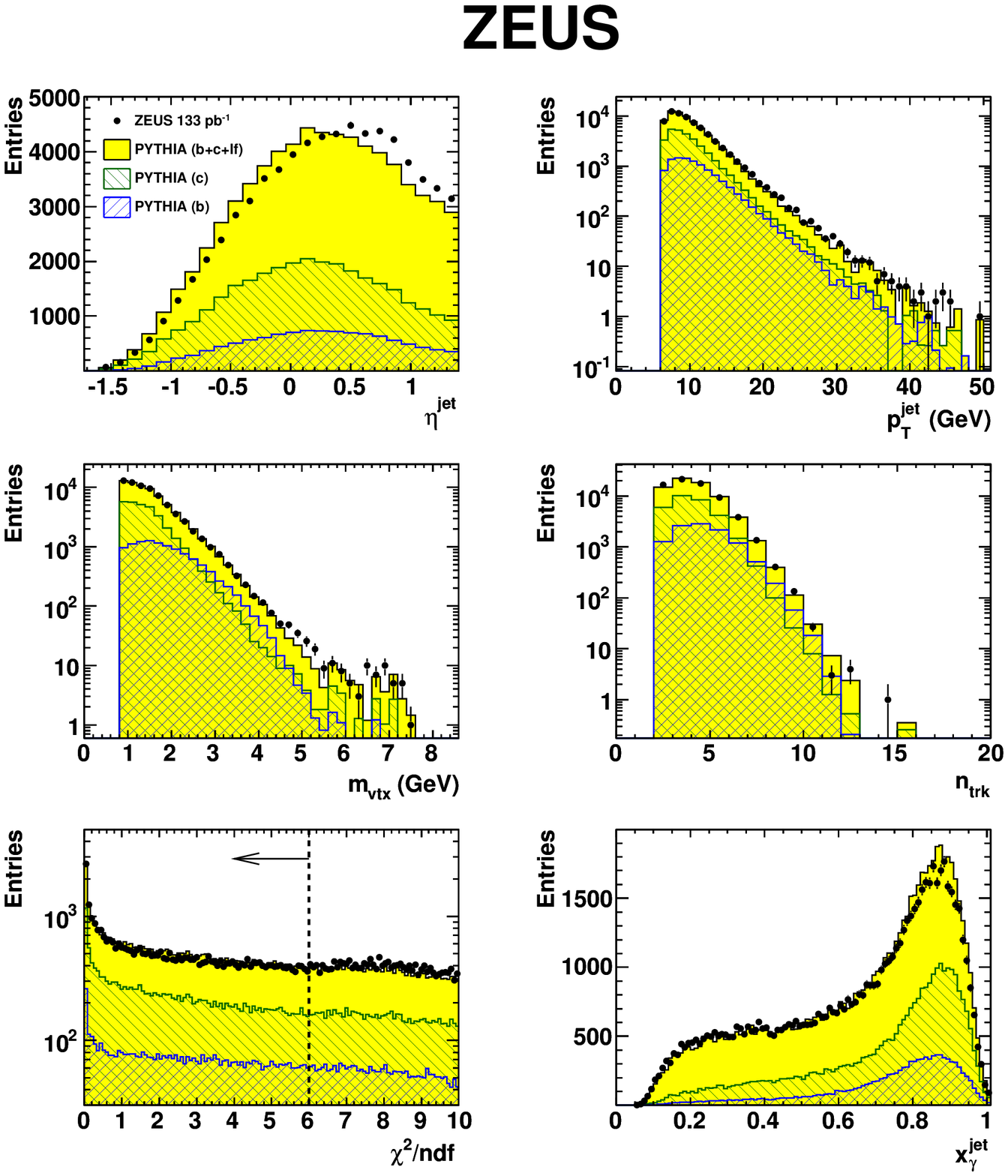} 
 \put(-9.05,14.7){\bf (a)}
 \put(-1.65,14.7){\bf (b)}
 \put(-9.05,9.5){\bf (c)}
 \put(-1.65,9.5){\bf (d)}
 \put(-9.05,4.3){\bf (e)}
 \put(-5.8,4.3){\bf (f)}
 \caption{Distributions of (a) \etajet and (b) \pTjet of the jets
   associated with a secondary vertex, (c) \mvtx and (d) \ntrk of the
   selected secondary vertices. (e) $\chi^{2}$/\ndf of the secondary
   vertices before the cut shown in the figure had been applied. (f)
   shows \xgamma weighted by the number of jets with associated
   secondary vertices in the event. The data are compared to the total
   MC distributions as well as the contributions from the beauty and
   charm MC subsamples. All samples were normalised according to the
   scaling factors obtained from the fit (see
   Section~\ref{bb-results}).}
\label{fig_cp}
\end{figure}

Figure~\ref{fig_cp} shows the data and MC distributions of \pTjet,
\etajet, \mvtx, the secondary vertex track multiplicity, \ntrk, and
$\chi^{2}$/\ndf of the secondary vertices. All distributions are shown
after all selection cuts, except for the $\chi^{2}$/\ndf distribution,
where the $\chi^{2}$/\ndf cut has not been applied yet. Also shown in
Fig.~\ref{fig_cp} is the fraction of the total hadronic $E-p_{Z}$
carried by the two highest-$p_{T}$ jets,
\begin{align*}
  \xgamma & = \frac{\sum _{j=1,2}(E^{j}-p^{j}_{Z})}{E-p_{Z}},\nonumber
\end{align*}
weighted by the number of jets with associated secondary vertices in
the event. This distribution is sensitive to the fraction of direct
and non-direct photoproduction contributions. The MC provides an
adequate description of the data for all variables except \etajet; the
effect of this discrepancy on the results is discussed in
Section~\ref{bb-syst}.

\begin{figure}[p]
\centering
 \includegraphics[scale=0.8]{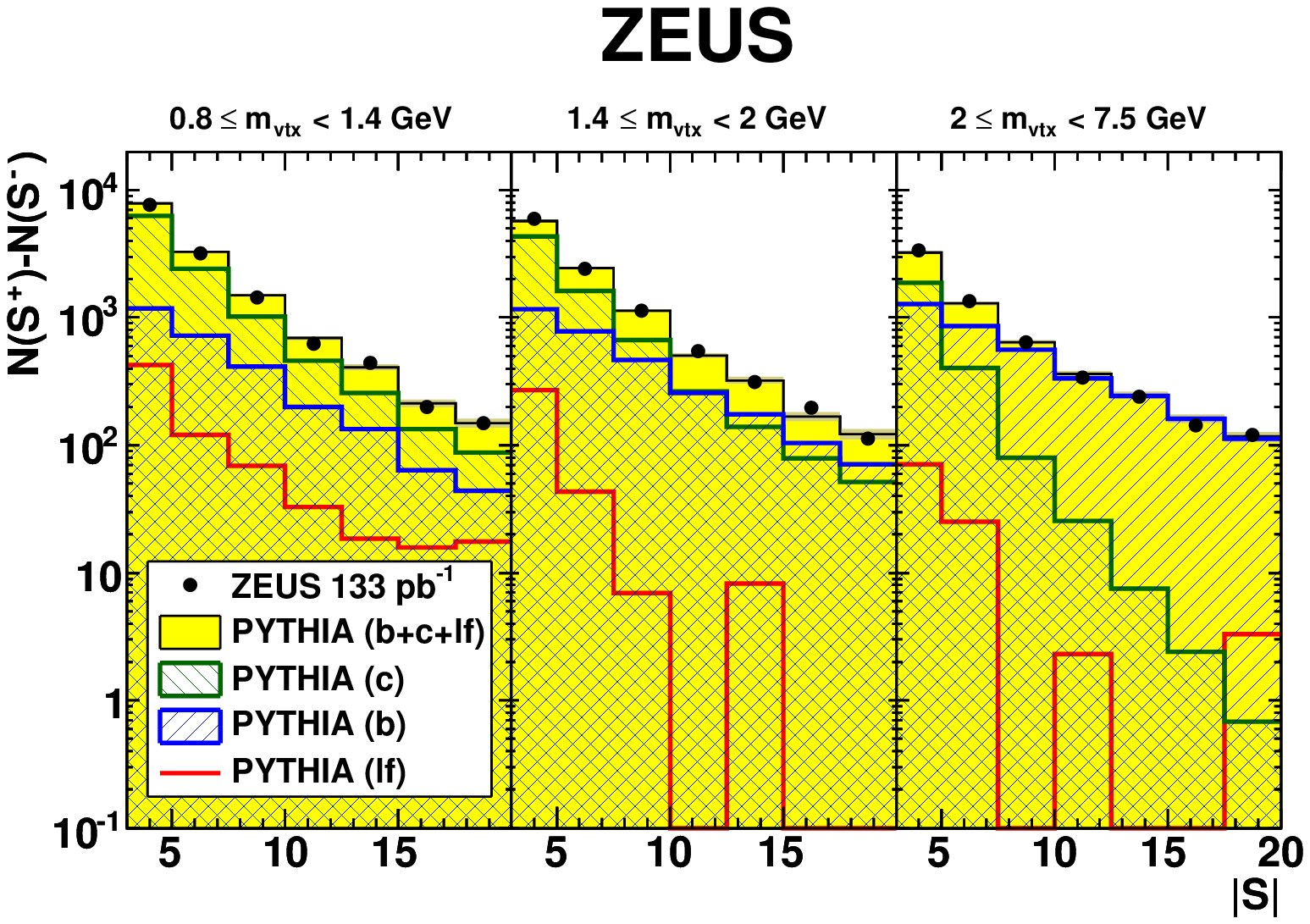}
 \caption{Distribution of the subtracted decay-length significance in
   three mass bins. The data are compared to the total PYTHIA MC
   distribution as well as the contributions from the beauty, charm
   and light-flavour MC subsamples. All samples were normalised
   according to the scaling factors obtained from the fit.}
\label{fig_subtracted}
\end{figure}

The beauty and charm contributions were extracted using a
least-squares fit\cite{thesis:ayagues:2008,thesis:vschoenb:2010} to
the subtracted distributions in the three mass bins. The MC beauty,
charm and light-flavour contributions, normalised to the data
luminosity, were scaled by the factors $k_{b}$, $k_{c}$ and
$k_{\text{lf}}$, respectively, to give the best fit to the observed
subtracted distributions. The overall MC normalisation was constrained
by requiring it to be consistent with the normalisation of the data in
the significance distribution with $|S|>3$ and $\unit[0.8]{} \le \mvtx
< \unit[7.5]{\Gev}$.  The subtracted and fitted distributions for the
three mass bins are shown in Fig.~\ref{fig_subtracted}.  The
contribution of the light flavours was substantially reduced through
the subtraction. After the subtraction, good agreement was also
observed between the data and the MC simulation. The fit procedure was
repeated in different bins of \pTjet and \etajet to obtain the
differential cross-sections \diffpt and \diffeta.

\begin{figure}[p]
\centering
  \includegraphics[scale=0.8]{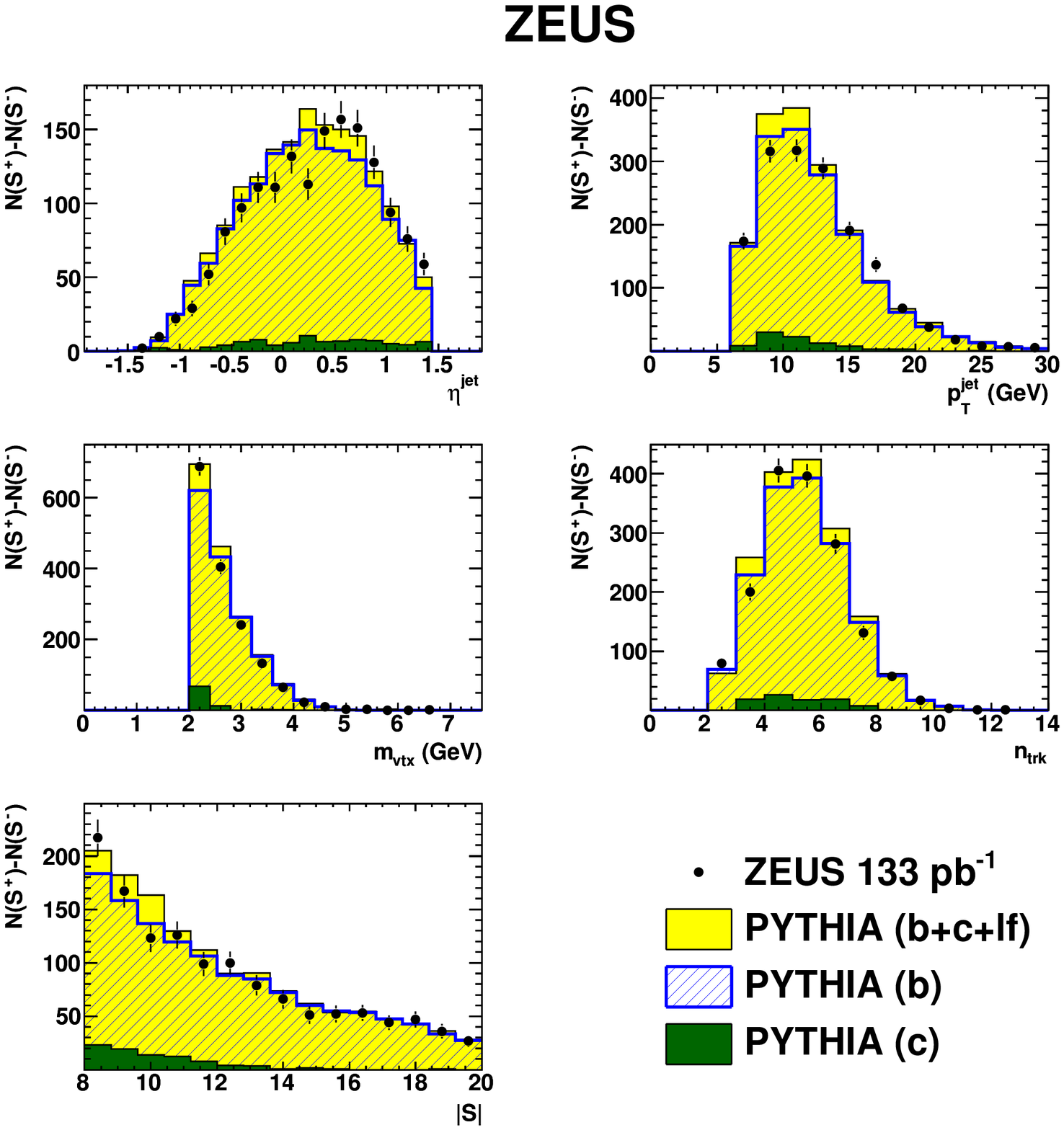}
  \put(-9.8,14.6){\bf (a)}
  \put(-1.8,14.6){\bf (b)}
  \put(-9.8,9.5){\bf (c)}
  \put(-1.8,9.5){\bf (d)}
  \put(-9.8,4.4){\bf (e)}
  \caption{Distributions of (a) \etajet, (b) \pTjet, (c) \mvtx and (d)
    \ntrk of the selected secondary vertices and (e) subtracted
    decay-length significance, for a beauty-enriched subsample with
    $\unit[2]{} \le \mvtx < \unit[7.5]{\gev}$ and $|S| > 8$. The data
    are compared to the total MC distributions as well as the
    contributions from the beauty and charm MC subsamples.  The
    light-flavour contribution is not shown separately as it is
    negligible on the scales shown.  All samples were normalised
    according to the scaling factors obtained from the fit.}
   \label{fig_enrichedcpb}
\end{figure}

\begin{figure}[p]
\centering
  \includegraphics[scale=0.8]{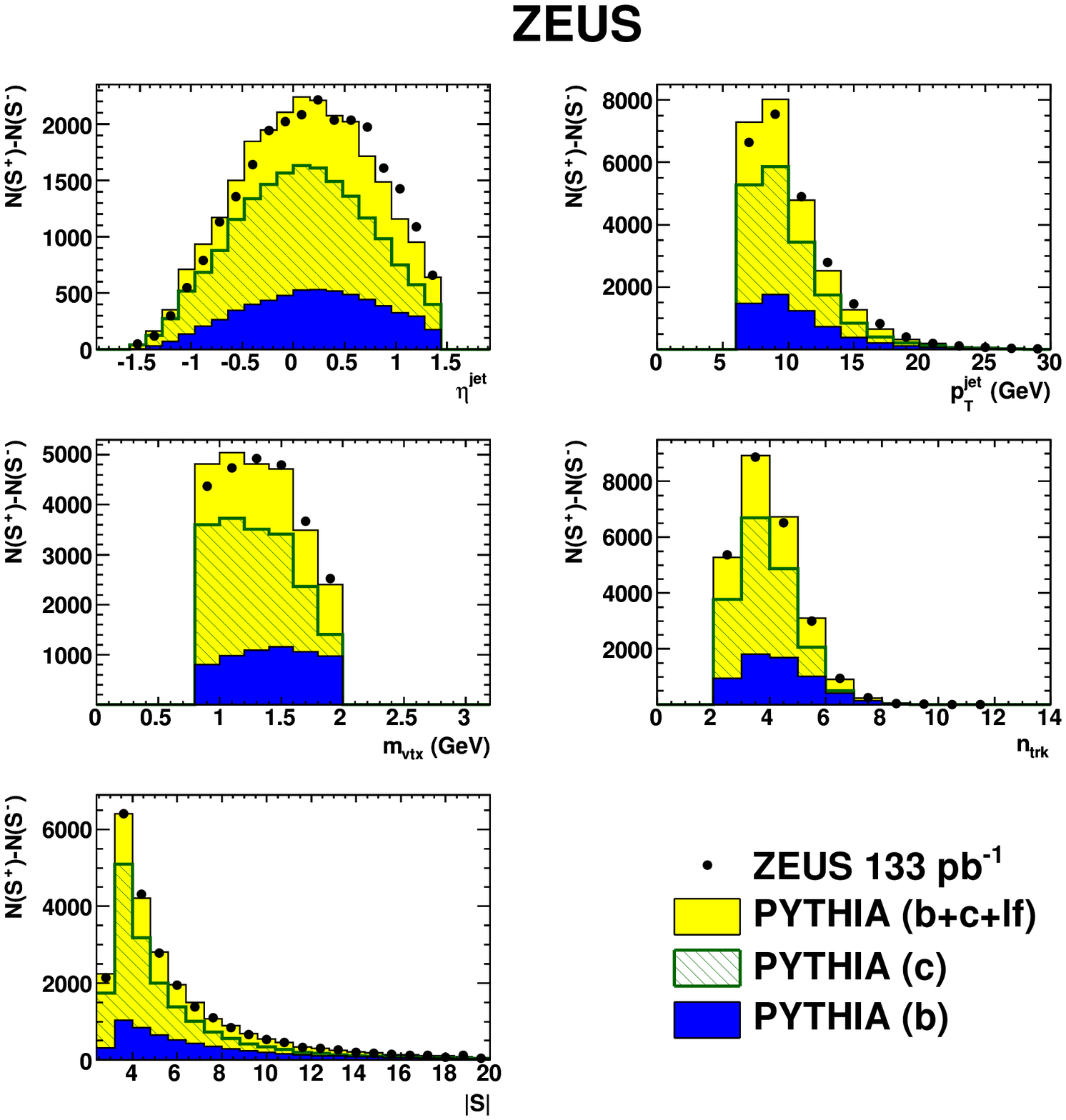}
  \put(-9.8,14.6){\bf (a)}
  \put(-1.8,14.6){\bf (b)}
  \put(-9.8,9.5){\bf (c)}
  \put(-1.8,9.5){\bf (d)}
  \put(-9.8,4.4){\bf (e)}
  \caption{Distributions of (a) \etajet, (b) \pTjet, (c) \mvtx and (d)
    \ntrk of the selected secondary vertices and (e) subtracted
    decay-length significance, for a charm-enriched subsample with
    $\unit[0.8]{} \le \mvtx < \unit[2]{\gev}$. No additional
    significance cut was applied here.  The data are compared to the
    total MC distributions as well as the contributions from the
    beauty and charm MC subsamples.  The light-flavour contribution is
    not shown separately as it is negligible on the scales shown.  All
    samples were normalised according to the scaling factors obtained
    from the fit.}
\label{fig_enrichedcpc}
\end{figure}

In order to check the quality of the data description by the MC,
subtracted distributions of \pTjet, \etajet, \mvtx, the
secondary-vertex track multiplicity, \ntrk, and $|S|$ are shown in
Fig.~\ref{fig_enrichedcpb} after beauty enrichment ($\unit[2]{} \le
\mvtx < \unit[7.5]{\Gev}$ and $|S| \ge 8$) and in
Fig.~\ref{fig_enrichedcpc} after charm enrichment ($\unit[0.8]{} \le
\mvtx < \unit[2]{\Gev}$).

The total visible cross section for inclusive heavy-quark jet
production, $\sigma^{q}$, with $q\in\{b,c\}$ is given by
\begin{equation*}\label{EqnXsecTot}
  \sigma^{q} = \frac{N_{q}^{\text{rec,Data}}}{\mathcal{A}_{q}\cdot\mathcal{L}_{\text{Data}}}.
\end{equation*}
Here, $\mathcal{L}_{\text{Data}}$ denotes the integrated luminosity,
$\mathcal{A}_{q}$ is the acceptance and $N_{q}^{\text{rec,Data}}$ the
number of reconstructed heavy-quark jets in data, which was determined
from the fit using
\begin{equation*}
N_{q}^{\text{rec,Data}} = k_{q}\cdot N_{q}^{\text{rec,MC}}\,,
\end{equation*}
with $N_{q}^{\text{rec,MC}}$ being the number of reconstructed events
in a MC sample with the same integrated luminosity as the
data. $k_{q}$ denotes the heavy-quark scaling factor obtained from the
fit. Defining the acceptance as
\begin{equation*}
\mathcal{A}_{q} = \frac{N_{q}^{\text{rec,MC}}}{N_{q}^{\text{true,HL}}}\,,
\end{equation*}
the cross section can be written as
\begin{equation*}
  \sigma^{q} = \frac{k_{q}\cdot N_{q}^{\text{true,HL}}}{\mathcal{L}^{\text{Data}}}.
\end{equation*}
Here, $N_{q}^{\text{true,HL}}$ denotes the number of generated heavy-quark
jets at hadron level (HL). Hadron-level jets were obtained by running
the $k_{T}$ clustering algorithm in the same mode as for the data with
the $E$-recombination scheme.  The algorithm was run on all
final-state MC particles before the decay of the weakly decaying $b$
or $c$ hadrons. True $b$ or $c$ jets were then defined as all
hadron-level jets containing a $b$ or $c$ hadron.  Signatures with $b$
or $c$ hadrons resulting from final-state gluon splitting ($g
\rightarrow q\bar q$) were also included in the respective signal,
independent of the quark flavours in the hard subprocess. The
contribution of gluon splitting to the beauty signal amounted to
$\approx\unit[2]{\%}$, while the contribution to the charm signal was
$\approx\unit[10]{\%}$.

The single-differential heavy-quark jet production cross section as a
function of a given variable, $v$, is defined accordingly:
\begin{align*}
  \frac{\dif\sigma^{q}}{\dif v} & =\frac{k_{q}\cdot
    N_{q}^{\text{true,HL}}}{\mathcal{L}^{\text{Data}}\cdot \Delta v}\, ,
\end{align*}
where $\Delta v$ is the width of the bin.

% ----------------------------------------------------------------------------
%   Systematic Uncertainties
% ----------------------------------------------------------------------------
\section{Systematic uncertainties}
\label{bb-syst}

Systematic uncertainties were evaluated by appropriate
variations of the MC
simulation. The fit of the subtracted decay-length
significance in \mvtx bins was repeated and the cross sections were
recalculated. The uncertainties on the total cross sections determined
for each source are summarised in Table~\ref{tab:syst}.
The following sources of experimental systematic uncertainties were
identified\cite{thesis:vschoenb:2010}:

\begin{enumerate}
\item the systematic uncertainties associated with the TLT and FLT
  trigger efficiency corrections (see
  Section~\ref{data-sel}) were determined by varying each
  correction within its estimated
  uncertainty;

\item the calorimetric part of the jet energy was 
  varied by $\unit[\pm 3]{\%}$;

\item the track-finding inefficiency in the data with respect to the
  MC was estimated to be at most \unit[2]{\%}. The overall
  uncertainty due to this tracking inefficiency was determined by
  randomly rejecting \unit[2]{\%} of all tracks in the MC and
  repeating the secondary vertex finding and all subsequent analysis steps;

\item the uncertainty due to the smearing procedure was estimated by
  varying the fraction of secondary vertices for which the decay
  length was smeared by $\unit[\pm 2]{\%}$. For variations of the
  fraction in this range the agreement between data and MC remained
  reasonable;

\item the uncertainty due to the asymmetry of the light-flavour
  content of the sample was evaluated by varying $k_{\text{lf}}$ by
  $\unit[\pm 11]{\%}$. The size of the variation was estimated from
  the uncertainty on the light-flavour fraction as determined by a fit
  to the subtracted decay-length significance distribution, where the
  overall normalisation constraint using the unsubtracted distribution
  was not applied;

\item the MC distributions for both light and heavy flavours were
  reweighted in \etajet and \pTjet to account for the differences
  between data and MC (see Fig.~\ref{fig_cp}). A
  reweighting of only the light-flavour content was also investigated.
  No significant change of the
  cross sections was observed and therefore no additional systematic
  uncertainty was assigned;

\item the various $D$ mesons have different lifetimes and
  decay modes. In order to account for the uncertainty of the
  different fragmentation fractions,
  the $D^{+}/D^{0}$ and $D^{+}/D^{+}_{s}$ ratios were varied by
  $\unit[\pm 10]{\%}$ while keeping the total number of $c$ hadrons
  constant;

\item the charm fragmentation function was varied by weighting all
  events according to
  \begin{displaymath}
    z = \frac{(E+P_{||})_{D}}{(E+P)_{c\text{-quark jet}}}
  \end{displaymath} 
  calculated in the string
  rest-frame\cite{cpc:135:238,*epj:c17:137,*hep-ph-0108264} such that
  the change in the mean value of $z$ corresponded to the measured
  uncertainty\cite{jhep:04-082};

\item the beauty fragmentation function was varied in analogy to the
  charm case using a variation of $z$ corresponding to a variation of
  the Peterson fragmentation parameter,
  $\varepsilon_{b}$, of $\pm 0.0015$~\cite{pr:d27:105,np:b565:245};

\item a $\unit[1.8]{\%}$ overall normalisation uncertainty was
  associated with the luminosity measurement. It was included in the
  systematic error on the total cross sections, but not in those of
  the differential cross sections.
\end{enumerate}

The same variations were applied to each bin for the differential
cross sections.
The total systematic uncertainty was obtained by adding the above
contributions in quadrature. In the case of beauty, the dominant effects
arise from the variation of the trigger-efficiency corrections, the
track-finding efficiency and the reweighting as a function of \pTjet.
For charm, the variation of the trigger-efficiency
corrections as well as the energy-scale variation contribute most to
the total systematic uncertainty.

As an additional consistency check, the contributions of direct and
non-direct photon processes were investigated by reweighting the
\xgamma distributions based on MC and data comparisons of the $b$- and
$c$-enriched samples. The effect on the cross sections was smaller
than that due to the reweighting of the \pTjet and \etajet
distributions and so a further contribution was not added to the
systematic uncertainties. A reweighting of the \mvtx distribution was
also done in order to account for residual differences between the
data and the MC. Its effect on the cross sections was found to be
negligible.

% ----------------------------------------------------------------------------
%   Theoretical predictions and uncertainties
% ----------------------------------------------------------------------------
\section{Theoretical predictions and uncertainties}
\label{bb-theo}

The measured total and differential cross sections were compared to
NLO QCD predictions calculated with the FMNR
programme\cite{np:b412:225}.  This calculation is based on the the
fixed-flavour-number scheme, using three light flavours for the charm
predictions and four for beauty.  The PDFs were taken from
CTEQ6.6\cite{jhep:07:012} for the proton and GRV-G
HO\cite{pr:d46:1973} for the photon.  The heavy-quark masses (pole
masses) were set to $m_{b}=4.75\gev$ and $m_{c}=1.5\gev$.  The QCD
scale, $\Lambda_{\mathrm{QCD}}^{(5)}$, was set to 0.226\gev.  The
renormalisation scale, $\mu_{R}$, and the factorisation scale,
$\mu_{F}$, were chosen to be equal and set to
$\mu_{R}=\mu_{F}=\frac{1}{2} \sqrt{\hat{p}_{T}^{2}+m_{b(c)}^{2}}$,
where $\hat{p}_{T}$ is the average transverse momentum of the heavy
quarks. In order to ease the comparison with previous analyses, the
theoretical predictions were also made using the
CTEQ5M\cite{epj:c12:375u} proton PDFs.

For the systematic uncertainty on the theoretical prediction, the
masses and scales were varied separately and the effects of both
variations were added in quadrature. The masses were varied using the
values $m_{b} = 4.5$ and $5.0\gev$, $m_{c} = 1.3$ and $1.7\gev$; the scales were
varied using
$\mu_{R}=\mu_{F}=\frac{1}{4}\sqrt{\hat{p}_{T}^{2}+m_{b(c)}^{2}}$ and
$\sqrt{\hat{p}_{T}^{2}+m_{b(c)}^{2}}$. The resulting uncertainties on
the NLO QCD predictions for the total cross sections are
$+\unit[22]{\%}$ and $-\unit[15]{\%}$ for beauty and $+\unit[42]{\%}$
and $-\unit[21]{\%}$ for charm.

Parton-level jets were found by applying the $k_{T}$ clustering
algorithm to the generated partonic final state in the same mode as
for the hadron level in the MC (see Section~\ref{bb-extr}). The NLO
QCD predictions for parton-level jets were corrected for hadronisation
effects. A bin-by-bin procedure was used whereby $\dif\sigma =
\dif\sigma_{\mathrm{NLO}} \cdot C_{\text{had}}$, and
$\dif\sigma_{\mathrm{NLO}}$ is the cross section for partons in the
final state of the NLO calculation. The hadronisation-correction
factors, $C_{\text{had}}$, were obtained from the ratio of the
hadron-level to the parton-level MC jet cross section, where the
parton level is defined as being the result of the parton-showering
stage of the simulation. The correction factors are given in
Tables~\ref{tab_dcs1} and~\ref{tab_dcs2}; their uncertainty was
negligible in comparison to the other theoretical
uncertainties\cite{pr:d70:012008b}.

% ----------------------------------------------------------------------------
%    Results
% ----------------------------------------------------------------------------
\section{Results}
\label{bb-results}

The total and single-differential beauty- and charm-jet cross sections
were measured for the processes
\begin{align*}
e^{-}p & \rightarrow e^{-}\,b(\bar{b})\,X\\
e^{-}p & \rightarrow e^{-}\,c(\bar{c})\,X
\end{align*}
in events with
\begin{equation*}
Q^{2} < \unit[1]{\Gev^{2}},\quad
0.2 < y < 0.8,\quad
p_{T}^{\text{jet}\,1(2)}>\unit[7(6)]{\Gev},\quad
-2.5 < \eta^{\text{jet}\,1(2)} < 2.5\,.
\end{equation*}
Here $\eta^{\text{jet}\,1(2)}$ and $p_{T}^{\text{jet}\,1(2)}$ refer,
respectively, to the pseudorapidities and the transverse momenta of
the two jets in the event with the largest transverse momentum within
the range $|\etajet| < 2.5$.  The cross sections are measured for
those jets among these two satisfying
\begin{equation*}
  -1.6 < \etaqjet < 1.4\,,
\end{equation*}
with $q \in \{b,c\}$.

The total beauty- and charm-jet production cross sections were
measured as
\begin{align*}
  \sigma^{\text{vis}}_{b} & = \unit[\phantom{5}682 \pm \phantom{1}21(\stat)^{+\phantom{5}52}_{-\phantom{5}52}(\syst)]{pb},\\
  \sigma^{\text{vis}}_{c} & = \unit[5780 \pm 120(\stat)^{+390}_{-410}(\syst)]{pb}.
\end{align*}
The errors given correspond to the statistical uncertainties and the
total systematic uncertainties including the errors due to the
uncertainty in the luminosity measurement. The measurements were
compared to NLO QCD predictions calculated with the FMNR programme
using the specifications given in Section~\ref{bb-theo}:
\begin{align*}
  \sigma^{\text{NLO}}_{b}\otimes C_{\text{had}}^{b} & = \unit[\phantom{5}740^{+\phantom{2}210}_{-\phantom{2}130}]{pb},\\
  \sigma^{\text{NLO}}_{c}\otimes C_{\text{had}}^{c} & = \unit[6000^{+2400}_{-1300}]{pb}.
\end{align*}
Hadronisation corrections of $C_{\text{had}}^{b} = \unit[0.84]{}$ and
$C_{\text{had}}^{c} = \unit[0.83]{}$ were applied to the NLO QCD
predictions. Good agreement between the measured cross sections and
the NLO QCD predictions is observed.  Replacing CTEQ6.6 by CTEQ5M as
proton PDF reduces the theory predictions by $\approx\unit[5]{\%}$.

\begin{figure}[p]
\centering
\hspace{0.5cm}\includegraphics[scale=0.77]{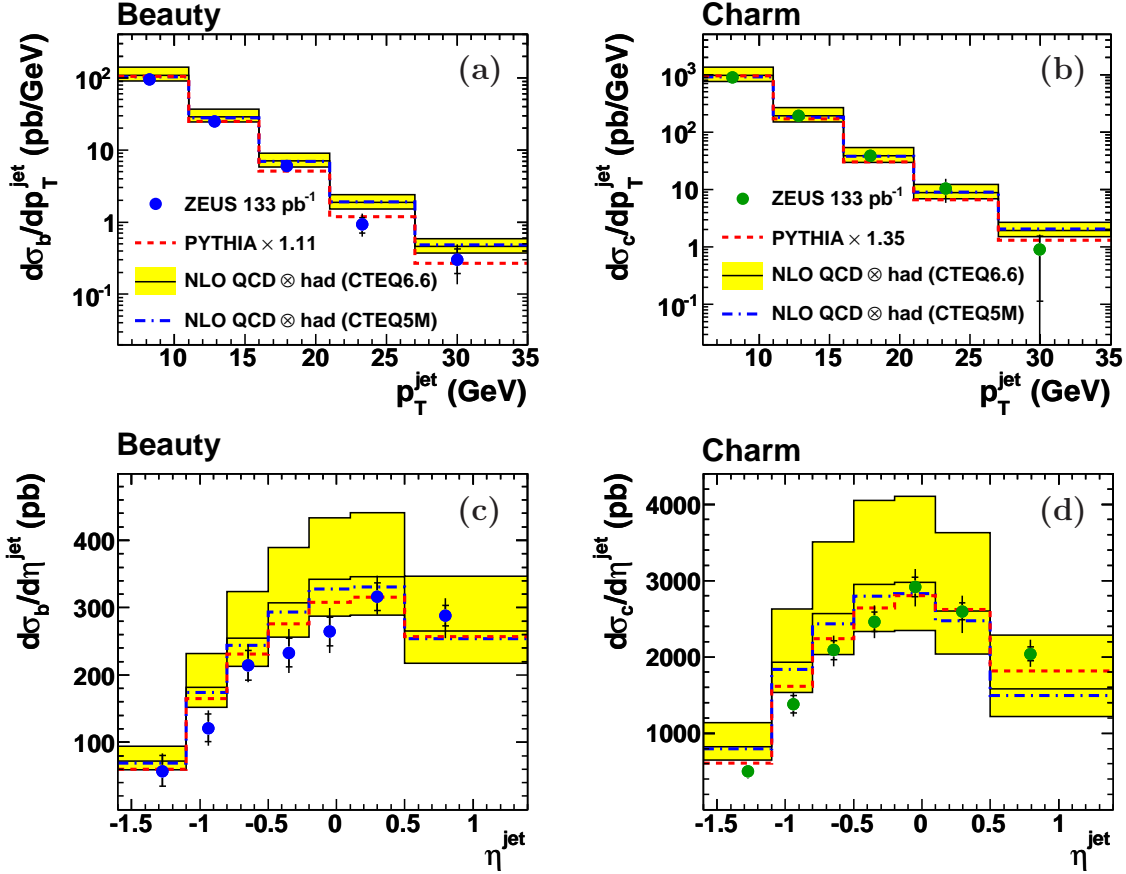}
\put(-9.4,11.4){\bf (a)}
\put(-1.75,11.4){\bf (b)}
\put(-9.4,5.55){\bf (c)}
\put(-1.75,5.55){\bf (d)}
\caption{Differential beauty-jet and charm-jet photoproduction cross
  sections as defined in Section~\ref{bb-results} as a function of
  (a)-(b) \pTjet and (c)-(d) \etajet. The data are shown as points.
  The inner error bars are the statistical uncertainties, while the outer
  error bars show the statistical and systematic uncertainties added
  in quadrature. The band represents the NLO QCD prediction, corrected
  for hadronisation effects, using CTEQ6.6 as proton PDF; the
  shaded band shows the estimated uncertainty. The NLO QCD prediction
  using CTEQ5M as proton PDF is depicted separately (dotted-dashed
  line). The scaled PYTHIA MC prediction (dashed line) is also shown.}
\label{fig_cs}
\end{figure}

The beauty and charm cross sections as a function of \pTjet and
\etajet are given in Tables~\ref{tab_dcs1} and~\ref{tab_dcs2},
respectively, and are shown in Fig.~\ref{fig_cs}. The measurements are
compared to the NLO QCD predictions and to the \PYTHIA MC scaled (see
Section~\ref{bb-extr}) by a factor of 1.11 for beauty and 1.35 for
charm, as obtained from the inclusive fit. The NLO QCD predictions are
in good agreement with the data and the scaled \PYTHIA MC describes
the distributions well.

\begin{figure}[p]
\centering
 \includegraphics[scale=0.6]{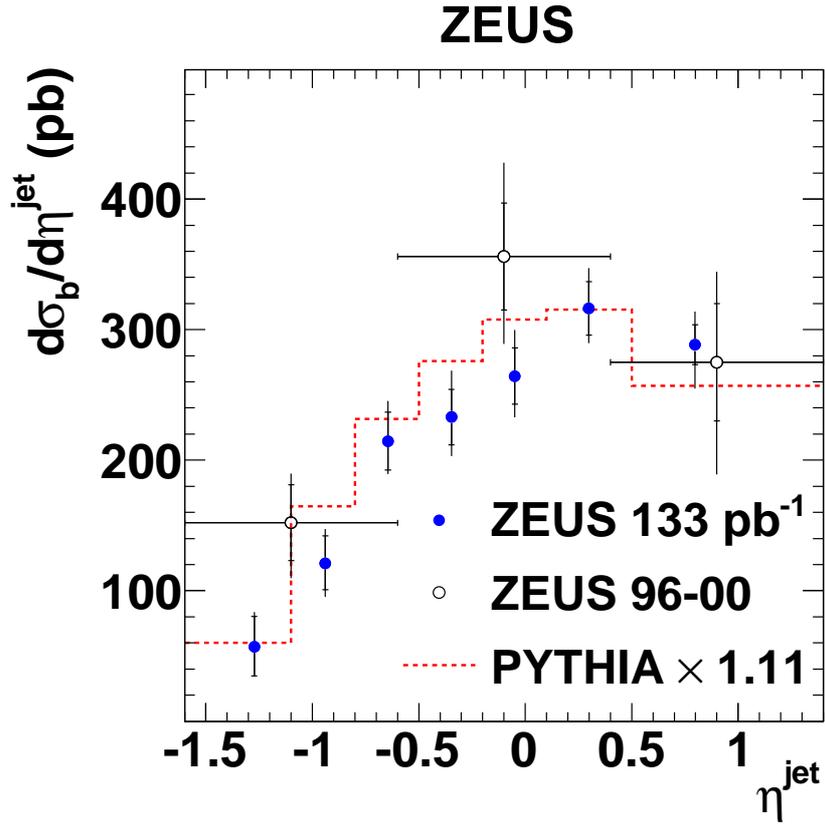}
 \caption{Differential beauty-jet photoproduction cross sections as a
   function of $\etajet$. The filled circles show the results from
   this analysis (the same data as shown in Fig.~\ref{fig_cs}(c)); the
   open circles show the results from a previously published
   measurement\protect\cite{pr:d70:012008b}. The inner error bars are
   the statistical uncertainties, while the outer error bars show the
   statistical and systematic uncertainties added in quadrature. The
   scaled PYTHIA MC prediction is also shown (dashed line).}
\label{fig_csmassimo}
\end{figure}

In Fig.~\ref{fig_csmassimo} the $b$-jet cross section, \diffeta, is
compared to a previously published analysis\cite{pr:d70:012008} using
semileptonic decays into muons in dijet events. Both measurements
agree well. The improved precision of this analysis can be clearly
seen. While a direct comparison with a previous H1 measurement using a
similar approach\cite{epj:c47:597} is not possible, as the
cross-section definitions are different, the relative errors on the
measurements in this paper are approximately a factor 3 (2) smaller
for beauty (charm).

\begin{figure}[p]
\centering
 \includegraphics[scale=0.8]{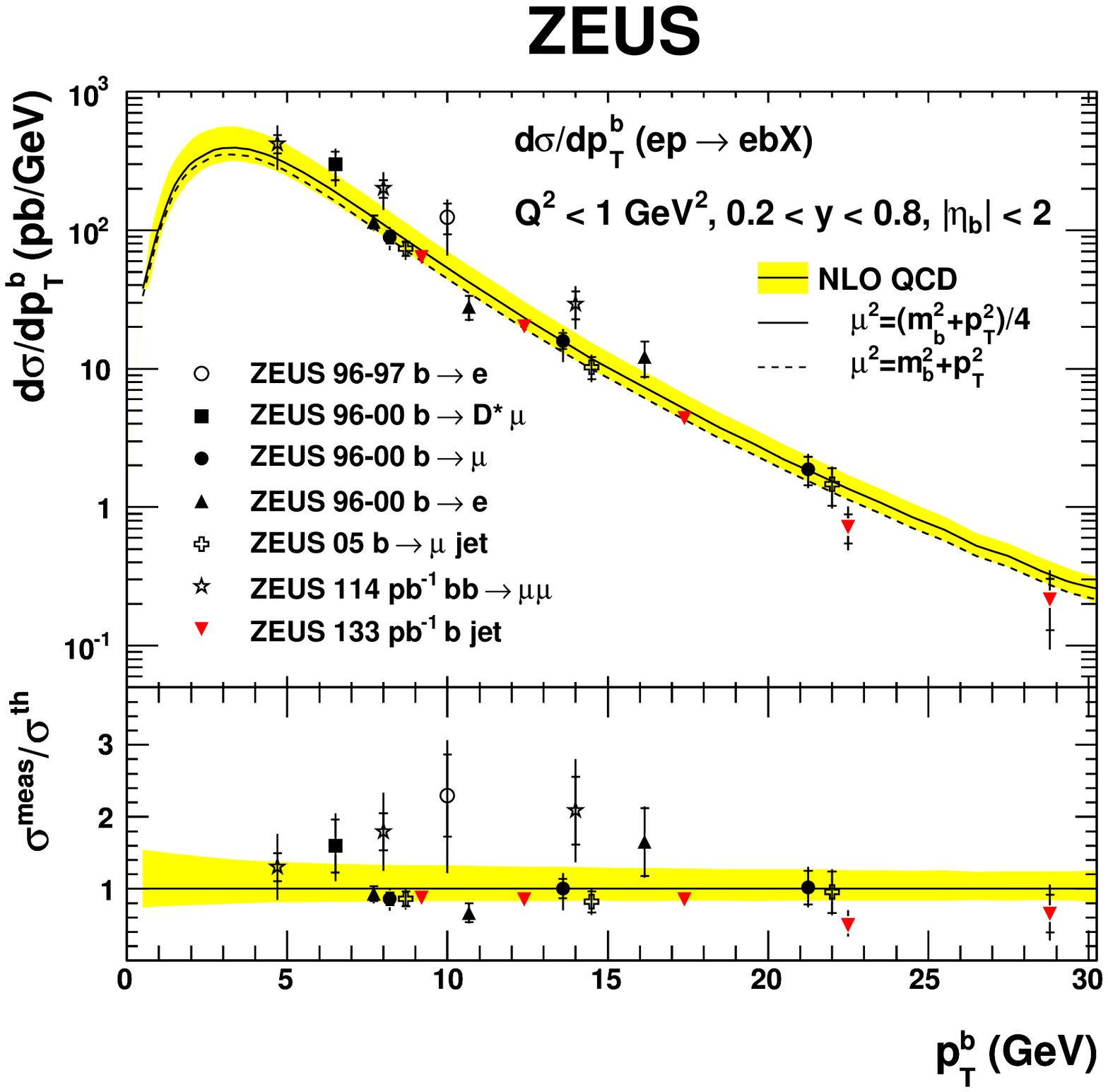}
 \put(-2.7,12.8){\bf (a)}
 \put(-2.7,5.3){\bf (b)}
 \caption{(a) Summary of differential cross sections for $b$-quark
   production as a function of \pTb as measured by the ZEUS
   collaboration. The measurements are shown as points, with the
   results of this analysis shown as inverted triangles. The inner
   error bars are the statistical uncertainties, while the outer error
   bars show the statistical and systematic uncertainties added in
   quadrature. The band represents the NLO QCD prediction and its
   theoretical uncertainty. The solid line shows the prediction for
   $\mu^2=(m_{b}^{2}+p_{T}^{2})/4$, while the dashed line shows the
   prediction for $\mu^2=m_{b}^{2}+p_{T}^{2}$. (b) The ratio of the
   measured cross sections, $\sigma^{meas}$, to the theoretical
   prediction, $\sigma^{th}$, for $\mu^2=(m_{b}^{2}+p_{T}^{2})/4$.}
\label{fig_csptb}
\end{figure}

In order to enable direct comparisons with other ZEUS measurements
given at the $b$-quark
level\cite{epj:c18:625,pr:d78:072001a,pr:d70:012008b,Chekanov:2008zz,Chekanov:2008tx,epj:c50:299},
the NLO QCD prediction corrected for hadronisation was used to
extrapolate the dijet cross sections to inclusive $b$-quark cross
sections:
\begin{equation*}
  \frac{\dif\sigma}{\dif\pTb} = 
  \frac{\left(\frac{\dif\sigma}{\dif\pTjet}\right)^{\text{vis}}}%
  {\left(\frac{\dif\sigma}{\dif\pTjet}\right)^{\text{NLO}}}
  \cdot \left(\frac{\dif\sigma}{\dif\pTb}\right)^{\text{NLO}}.
\end{equation*}
For the previous measurements, the extrapolations have been updated
using the CTEQ6.6 proton PDFs. In Fig.~\ref{fig_csptb}, the $b$-quark
differential cross sections as a function of the quark transverse
momentum, $\dif\sigma(ep\rightarrow bX)/\dif p^{b}_{T}$, are shown for
$b$-quark pseudorapidity in the laboratory frame, $|\eta_{b}|<2$, for
$Q^{2}<\unit[1]{\Gev^{2}}$ and $0.2<y<0.8$. The $\bar{b}$ quark was
not taken into account in the definition of the $b$-quark cross
section. The measurement presented here extends the kinematic region
to higher \pTb values than previous measurements and represents the
most precise measurement of $b$-quark photoproduction at HERA. Good
agreement with the NLO QCD prediction is observed for many independent
ZEUS measurements, giving a consistent picture of $b$-quark
photoproduction over a wide range of \pTb.

\begin{figure}[p]
\centering
 \includegraphics[scale=0.8]{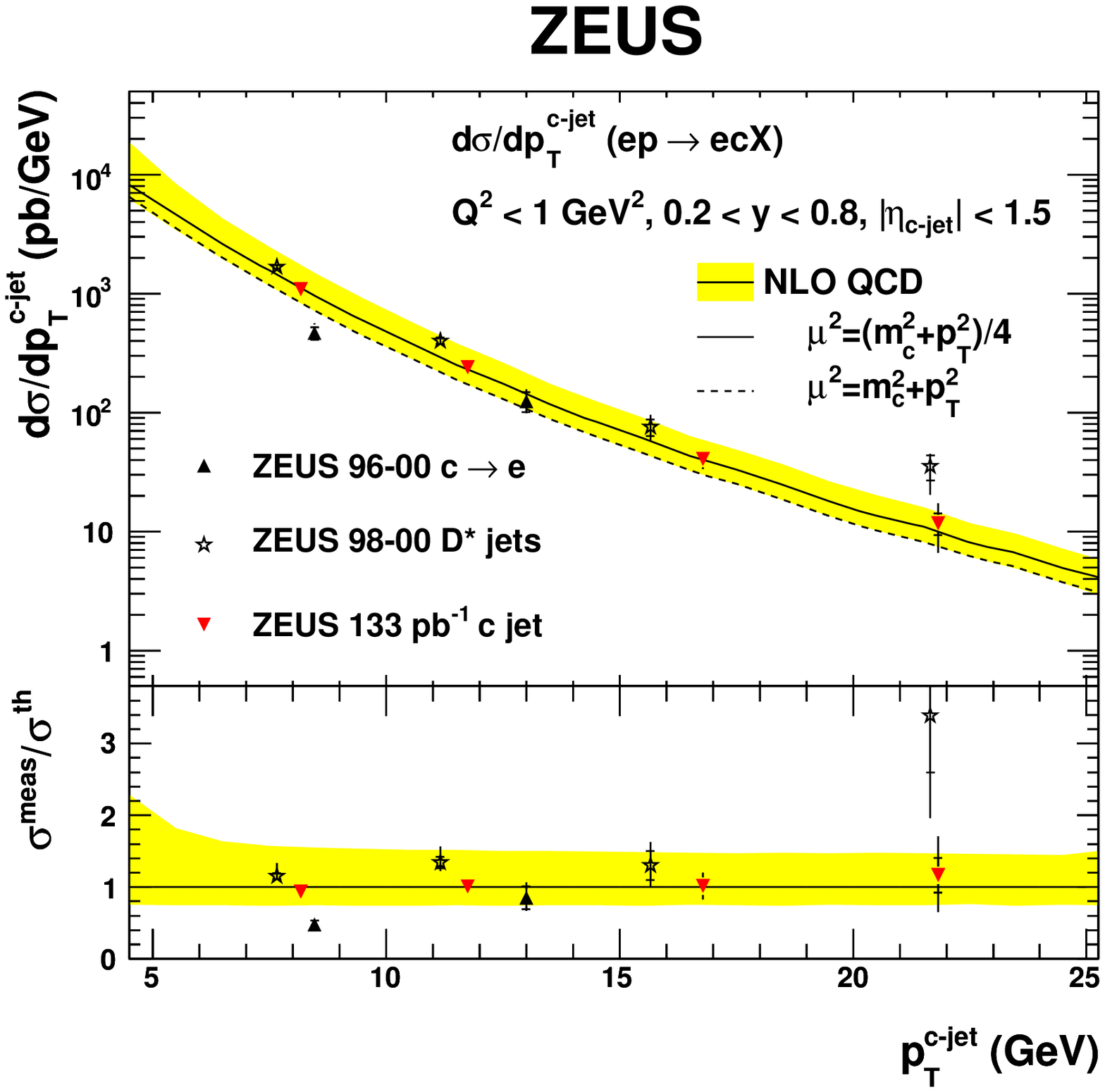}
 \put(-2.7,12.8){\bf (a)}
 \put(-2.7,5.3){\bf (b)}
 \caption{(a) Summary of differential cross sections for $c$-quark jet
   production as a function of \pTcjet (a) as measured by the ZEUS
   collaboration. The measurements are shown as points, with the
   results of this analysis shown as inverted triangles. The inner
   error bars are the statistical uncertainties, while the outer error
   bars show the statistical and systematic uncertainties added in
   quadrature. The band represents the NLO QCD prediction and its
   theoretical uncertainty. The solid line shows the prediction for
   $\mu^2=(m_{c}^{2}+p_{T}^{2})/4$, while the dashed line shows the
   prediction for $\mu^2=m_{c}^{2}+p_{T}^{2}$. (b) The ratio of the
   measured cross sections, $\sigma^{meas}$, to the theoretical
   prediction, $\sigma^{th}$, for $\mu^2=(m_{c}^{2}+p_{T}^{2})/4$.}
 \label{fig_csptc}
\end{figure}

The corresponding $c$-quark cross sections were also calculated and
are shown in Fig.~\ref{fig_csptc}. Due to the lower mass of the charm
quark, its momentum is more affected by gluon radiation. Hence the
corresponding cross section is shown as a function of the parton-level
jet momentum (calculated as in Section~\ref{bb-theo}) rather than that
of the quark. Here the cross sections have been extrapolated to the
region $|\eta_{c\text{-jet}}|<1.5$, as it corresponded better to the
measurements. 

The $c$-quark jet cross sections are consistent with previous
ZEUS measurements\cite{pr:d78:072001a,Chekanov2005492} and are in
good agreement with the NLO QCD prediction.

% ----------------------------------------------------------------------------
%    Conclusions
% ----------------------------------------------------------------------------
\section{Conclusions}
\label{bb-conclusions}

Inclusive beauty- and charm-jet cross sections in photoproduction at
HERA have been presented, exploiting the long lifetimes and large
masses of $b$ and $c$ hadrons.  Compared to previous measurements of
specific decay chains, this analysis has substantially increased
statistics and a reduced dependence on the branching fractions.  The
background from light-quark jets was suppressed by using the
subtracted decay-length significance distribution of secondary
vertices.

The visible cross sections as well as differential cross sections as a
function of \pTjet and \etajet have been compared with NLO QCD
calculations. Good agreement is observed.

In order to be able to compare these cross sections with others, they
have been extrapolated to the region $|\eta_{b}| < 2$
($|\eta_{c\text{-jet}}| < 1.5$) using the NLO QCD predictions. Cross
sections as a function of the transverse momentum of the $b$ quark and
of the $c$-quark jet have been determined and compared with previous
ZEUS measurements.  The measurements agree with each other and give a
consistent picture of heavy-quark photoproduction over a wide
kinematic range.

The charm cross sections presented in this paper are more precise than
previous measurements made by the ZEUS collaboration and have similar
accuracy as measurements made by H1. The beauty cross sections
represent the most precise measurements of $b$-quark photoproduction
made at HERA.

% ----------------------------------------------------------------------------
%       Mandatory acknowledgements. You may add your buddies to it.c
% ----------------------------------------------------------------------------
\section*{Acknowledgements}
\label{sec-ack}

We appreciate the contributions to the construction and maintenance of
the ZEUS detector of many people who are not listed as authors.  The
HERA machine group and the DESY computing staff are especially
acknowledged for their success in providing excellent operation of the
collider and the data-analysis environment. We thank the DESY
directorate for their strong support and encouragement.

\vfill\eject

%%% Local Variables: 
%%% mode: latex
%%% TeX-master: "DESY-11-067"
%%% End: 

%------------------------------------------------------------------------------
%       Bibliography
%------------------------------------------------------------------------------
\clearpage
{\raggedright
\providecommand{\etal}{et al.\xspace}
\providecommand{\coll}{Collab.\xspace}
\catcode`\@=11
\def\@bibitem#1{%
\ifmc@bstsupport
  \mc@iftail{#1}%
    {;\newline\ignorespaces}%
    {\ifmc@first\else.\fi\orig@bibitem{#1}}
  \mc@firstfalse
\else
  \mc@iftail{#1}%
    {\ignorespaces}%
    {\orig@bibitem{#1}}%
\fi}%
\catcode`\@=12

}
%\include{DESY-11-067-ref}
%(for submission to ZEUS sec and arxiv include bbl as bibtex will not be run)
%------------------------------------------------------------------------------
%       Tables
%------------------------------------------------------------------------------
%-------------------------------------------------------------------------------
%       Systematics
%-------------------------------------------------------------------------------

\begin{table}[h!]
  \centering
    \begin{tabular}{rl|c@{ / }c}
      \hline
      & Source & Beauty & Charm \\
      & & \multicolumn{2}{c}{(\%)} \\\hline
      1a) & TLT trigger efficiency & $\unit[\pm 0.8]{}$ & $\unit[\pm 2.0]{}$\\\hline
      1b) & FLT trigger efficiency & $^{\unit[+4.1]{}}_{\unit[-3.8]{}}$ & $^{\unit[+4.0]{}}_{\unit[-3.7]{}}$\\\hline
      2) & CAL hadronic energy scale & $\unit[\pm 0.6]{}$ & $\unit[\pm 4.3]{}$\\\hline
      3) & Track-finding uncertainty & $\unit[+5.9]{}$ & $\unit[+1.0]{}$\\\hline
      4) & Decay-length smearing & $\unit[\pm 1.0]{}$ & $\unit[\pm 0.7]{}$\\\hline
      5) & Light-flavour asymmetry & ${\unit[\pm 0.2]{}}$ & ${\unit[\pm 0.7]{}}$\\\hline
      6a) & \etajet reweighting & $\unit[-1.2]{}$ & $\unit[-1.0]{}$\\\hline
      6b) & \pTjet reweighting & $\unit[-5.5]{}$ & $\unit[-1.1]{}$\\\hline
      7a) & $D^{\pm}/D^{0}$ ratio & $^{\unit[+0\phantom{.0}]{}}_{\unit[-1.3]{}}$ & $^{\unit[+0.6]{}}_{\unit[-1.8]{}}$\\\hline
      7b) & $D^{\pm}/D^{\pm}_{s}$ ratio & $^{\unit[+0\phantom{.0}]{}}_{\unit[-1.2]{}}$ & $^{\unit[+0.1]{}}_{\unit[-1.3]{}}$\\\hline
      8) & Charm fragmentation & $^{\unit[+0.3]{}}_{\unit[-0.3]{}}$ & $^{\unit[+1.2]{}}_{\unit[-1.3]{}}$\\\hline
      9) & Beauty fragmentation & $^{\unit[+1.8]{}}_{\unit[-2.1]{}}$ & $^{\unit[+0.1]{}}_{\unit[-0.1]{}}$\\\hline
      10) & Luminosity measurement & $\unit[\pm 1.8]{}$ & $\unit[\pm 1.8]{}$\\\hline\hline
      & Total & $^{\unit[+7.8]{}}_{\unit[-7.7]{}}$ & $^{\unit[+6.7]{}}_{\unit[-7.0]{}}$\\\hline
    \end{tabular}
    \caption[Systematic uncertainties]{Systematic
      uncertainties on the total beauty- and charm-jet cross sections.}
    \label{tab:syst}
\end{table}

%-------------------------------------------------------------------------------
%       Cross sections
%-------------------------------------------------------------------------------

\begin{table}[h!]
  \begin{center}
    \begin{tabular}{r@{ : }l|c|c|c}
      \hline
      \multicolumn{2}{c|}{\pTjet} & \diffptb & \diffnloptb & \cbhad \\
      \multicolumn{2}{c|}{(GeV)} & (pb/GeV) & (pb/GeV) & \\\hline
      $\phantom{1}$6 & 11 & $\unit[95.6\phantom{2}\pm 4.9\phantom{4}^{+9.8\phantom{2}}_{-7.0\phantom{2}}]{}$ & 
      $\unit[109\phantom{.22}^{+31\phantom{.22}}_{-19\phantom{.22}}]{}$ & 0.83 \\
      11 & 16 & $\unit[24.8\phantom{2}\pm 1.2\phantom{4}^{+1.8\phantom{2}}_{-1.4\phantom{2}}]{}$ & 
      $\unit[\phantom{1}29.1\phantom{2}^{+\phantom{1}7.9\phantom{2}}_{-\phantom{1}4.7\phantom{2}}]{}$ & 0.89 \\
      16 & 21 & $\unit[\phantom{2}6.02\pm 0.49^{+0.55}_{-0.57}]{}$ & 
      $\unit[\phantom{10}7.1\phantom{2}^{+\phantom{1}2.0\phantom{2}}_{-\phantom{1}1.2\phantom{2}}]{}$ & 0.92 \\
      21 & 27 & $\unit[\phantom{2}0.93\pm 0.22^{+0.31}_{-0.20}]{}$ & 
      $\unit[\phantom{10}1.87^{+\phantom{1}0.54}_{-\phantom{1}0.34}]{}$ & 0.95 \\
      27 & 35 & $\unit[\phantom{2}0.30\pm 0.12^{+0.14}_{-0.12}]{}$ & 
      $\unit[\phantom{10}0.46^{+\phantom{1}0.13}_{-\phantom{1}0.08}]{}$ & 1.05 \\\hline
      \multicolumn{5}{c}{}\\\hline
      \multicolumn{2}{c|}{\etajet} & \diffetab & \diffnloetab & \cbhad \\
      \multicolumn{2}{c|}{} & (pb) & (pb) & \\\hline
       $\unit[-1.6]{}$ & $\unit[-1.1]{}$ & $\unit[\phantom{2}57\pm 22^{+13}_{-\phantom{1}3}]{}$ & 
       $\unit[\phantom{1}72^{+22}_{-13}]{}$ & 0.70 \\
       $\unit[-1.1]{}$ & $\unit[-0.8]{}$ & $\unit[121\pm 21^{+16}_{-16}]{}$ & 
       $\unit[182^{+50}_{-30}]{}$ & 0.78 \\
       $\unit[-0.8]{}$ & $\unit[-0.5]{}$ & $\unit[214\pm 22^{+22}_{-12}]{}$ & 
       $\unit[255^{+69}_{-42}]{}$ & 0.79 \\
       $\unit[-0.5]{}$ & $\unit[-0.2]{}$ & $\unit[233\pm 21^{+28}_{-21}]{}$ & 
       $\unit[307^{+83}_{-50}]{}$ & 0.79 \\
       $\unit[-0.2]{}$ & $\unit[\phantom{-}0.1]{}$ & $\unit[264\pm 22^{+28}_{-23}]{}$ & 
       $\unit[342^{+91}_{-55}]{}$ & 0.81 \\
       $\unit[\phantom{-}0.1]{}$ & $\unit[\phantom{-}0.5]{}$ & $\unit[316\pm 21^{+23}_{-17}]{}$ & 
       $\unit[346^{+96}_{-57}]{}$ & 0.86 \\
       $\unit[\phantom{-}0.5]{}$ & $\unit[\phantom{-}1.4]{}$ & $\unit[288\pm 15^{+20}_{-30}]{}$ & 
       $\unit[265^{+82}_{-48}]{}$ & 0.93 \\\hline
      \multicolumn{5}{c}{}
    \end{tabular}
    \caption[Beauty-jet production cross sections and NLO
    predictions]{Summary table of differential beauty-jet photoproduction
      cross sections, as defined in Section~\ref{bb-results}.
      The measurements are given together with their statistical
      and systematic uncertainties.  The NLO QCD predictions using
      CTEQ6.6 and their uncertainty are also listed. The last column
      gives the hadronisation correction factors, \cbhad.  }
    \label{tab_dcs1}
  \end{center}
\end{table}

\begin{table}[h!]
  \begin{center}
    \begin{tabular}{r@{ : }l|c|c|c}
      \hline
      \multicolumn{2}{c|}{\pTjet} & \diffptc & \diffnloptc & \cchad \\
      \multicolumn{2}{c|}{(\gev)} & (pb/\gev) & (pb/\gev) & \\\hline
      $\phantom{1}$6 & 11 & $\unit[906\phantom{.2}\pm 24\phantom{.2}^{+56\phantom{.2}}_{-60\phantom{.2}}]{}$ &
      $\unit[967\phantom{.22}^{+380\phantom{.22}}_{-210\phantom{.22}}]{}$ & 0.82 \\
      11 & 16 & $\unit[194\phantom{.2}\pm \phantom{1}7\phantom{.2}^{+20\phantom{.2}}_{-20\phantom{.2}}]{}$ & 
      $\unit[192\phantom{.22}^{+\phantom{1}75\phantom{.22}}_{-\phantom{1}41\phantom{.22}}]{}$ & 0.90 \\
      16 & 21 & $\unit[\phantom{1}39.1\pm \phantom{1}3.3^{+\phantom{2}6.4}_{-\phantom{2}6.4}]{}$ & 
      $\unit[\phantom{1}38.5\phantom{2}^{+\phantom{1}15\phantom{.22}}_{-\phantom{10}8.5\phantom{2}}]{}$ & 0.92 \\
      21 & 27 & $\unit[\phantom{1}10.5\pm \phantom{1}2.1^{+\phantom{2}4.4}_{-\phantom{2}4.0}]{}$ & 
      $\unit[\phantom{12}8.9\phantom{2}^{+\phantom{12}3.4\phantom{2}}_{-\phantom{12}2.0\phantom{2}}]{}$ & 0.90 \\
      27 & 35 & $\unit[\phantom{12}0.9\pm \phantom{1}0.7^{+\phantom{2}0.4}_{-\phantom{2}0.9}]{}$ & 
      $\unit[\phantom{12}1.96^{+\phantom{12}0.72}_{-\phantom{12}0.43}]{}$ & 0.91 \\\hline
      \multicolumn{5}{c}{}\\\hline
      \multicolumn{2}{c|}{\etajet} & \diffetac & \diffnloetac & \cchad \\
      \multicolumn{2}{c|}{} & (pb) & (pb) & \\\hline
       $\unit[-1.6]{}$ & $\unit[-1.1]{}$ & $\unit[\phantom{2}499\pm\phantom{1}79^{+\phantom{1}36}_{-\phantom{1}46}]{}$ & 
       $\unit[\phantom{1}825^{+\phantom{1}320}_{-\phantom{1}180}]{}$ & 0.71 \\
       $\unit[-1.1]{}$ & $\unit[-0.8]{}$ & $\unit[1380\pm 110^{+110}_{-110}]{}$ & 
       $\unit[1933^{+\phantom{1}700}_{-\phantom{1}400}]{}$ & 0.79\\
       $\unit[-0.8]{}$ & $\unit[-0.5]{}$ & $\unit[2090\pm 120^{+140}_{-180}]{}$ & 
       $\unit[2566^{+\phantom{1}940}_{-\phantom{1}540}]{}$ & 0.80\\
       $\unit[-0.5]{}$ & $\unit[-0.2]{}$ & $\unit[2460\pm 130^{+170}_{-170}]{}$ & 
       $\unit[2948^{+1100}_{-\phantom{1}610}]{}$ & 0.80\\
       $\unit[-0.2]{}$ & $\unit[\phantom{-}0.1]{}$ & $\unit[2920\pm 130^{+200}_{-220}]{}$ & 
       $\unit[2975^{+1100}_{-\phantom{1}630}]{}$ & 0.83\\
       $\unit[\phantom{-}0.1]{}$ & $\unit[\phantom{-}0.5]{}$ & $\unit[2600\pm 110^{+180}_{-260}]{}$ & 
       $\unit[2602^{+1000}_{-\phantom{1}560}]{}$ & 0.87\\
       $\unit[\phantom{-}0.5]{}$ & $\unit[\phantom{-}1.4]{}$ & $\unit[2040\pm\phantom{1}91^{+160}_{-140}]{}$ & 
       $\unit[1579^{+\phantom{1}700}_{-\phantom{1}360}]{}$ & 0.89\\\hline
      \multicolumn{5}{c}{}
    \end{tabular}
    \caption[Charm-jet production cross sections and NLO
    predictions]{Summary table of differential charm-jet photoproduction
      cross sections, as defined in Section~\ref{bb-results}.
      The measurements are given together with their statistical
      and systematic uncertainties.  The NLO QCD predictions using
      CTEQ6.6 and their uncertainty are also listed. The last column
      gives the hadronisation correction factors, \cchad.  }
    \label{tab_dcs2}
  \end{center}
\end{table}

%%% Local Variables: 
%%% mode: latex
%%% TeX-master: "DESY-11-067"
%%% End: 

%------------------------------------------------------------------------------
%       Figures
%------------------------------------------------------------------------------
%-------------------------------------------------------------------------------
%       Results
%-------------------------------------------------------------------------------

%%% Local Variables: 
%%% mode: latex
%%% TeX-master: "DESY-11-067"
%%% End: 

%
%       ... that's it
%
\end{document}